%% file: main.tex
\documentclass[aps,prb,imp,superscriptaddress,longbibliography]{revtex4}
\usepackage[pdftex,
 letterpaper=true,
 hyperindex=true,
 breaklinks=true,
 colorlinks=false,
 citecolor=blue,
 pdftitle={},
 pdfauthor={}]
{hyperref}
\usepackage{natbib}

\DeclareUnicodeCharacter{2215}{/}

\usepackage[normalem]{ulem}

\usepackage{graphicx}
\usepackage{xcolor} 
\usepackage{setspace}
\usepackage[T1]{fontenc}
\usepackage{amsmath}
\usepackage{mathtools}
\usepackage{amssymb}
\usepackage{amsthm}

\usepackage[]{multibib}
\newcites{M}{methods}

\begin{document}

\title{Extreme Shape Coexistence observed in $^{70}$Co}

\author{Cade Dembski}
    \email[]{cdembski@nd.edu}
	\affiliation{Facility for Rare Isotope Beams, Michigan State University, East Lansing, Michigan, 48824, USA}
	\affiliation{Department of Physics and Astronomy, University of Notre Dame, Notre Dame, Indiana, 46556, USA}

\author{Artemis Spyrou}
    \email[]{spyrou@frib.msu.edu}
	\affiliation{Facility for Rare Isotope Beams, Michigan State University, East Lansing, Michigan, 48824, USA}
	\affiliation{Department of Physics \& Astronomy, Michigan State University, East Lansing, Michigan, 48824, USA}

 \author{B. Alex Brown}
 	\affiliation{Facility for Rare Isotope Beams, Michigan State University, East Lansing, Michigan, 48824, USA}
	\affiliation{Department of Physics \& Astronomy, Michigan State University, East Lansing, Michigan, 48824, USA}

\author{Sean N. Liddick}
	\affiliation{Facility for Rare Isotope Beams, Michigan State University, East Lansing, Michigan, 48824, USA}
	\affiliation{Department of Chemistry, Michigan State University, East Lansing, Michigan, 48824, USA}

\author{Hannah C. Berg}
    \affiliation{Facility for Rare Isotope Beams, Michigan State University, East Lansing, Michigan, 48824, USA}
	\affiliation{Department of Physics \& Astronomy, Michigan State University, East Lansing, Michigan, 48824, USA}

\author{Darren L. Bleuel}
    \affiliation{Lawrence Livermore National Laboratory, Livermore, California, 94551, USA}

\author{Katherine Childers}
    \affiliation{Facility for Rare Isotope Beams, Michigan State University, East Lansing, Michigan, 48824, USA}
	\affiliation{Department of Chemistry, Michigan State University, East Lansing, Michigan, 48824, USA}

\author{Benjamin P. Crider}
    \affiliation{Department of Physics and Astronomy, Mississippi State University, Mississippi State, Mississippi, 39762, USA}

\author{Alexander C. Dombos}
    \affiliation{Facility for Rare Isotope Beams, Michigan State University, East Lansing, Michigan, 48824, USA}
	\affiliation{Department of Physics \& Astronomy, Michigan State University, East Lansing, Michigan, 48824, USA}

\author{Erin C. Good}
    \affiliation{Pacific Northwest National Laboratory, Richland, Washington, 99354, USA}

\author{Caley Harris}
    \affiliation{Facility for Rare Isotope Beams, Michigan State University, East Lansing, Michigan, 48824, USA}
	\affiliation{Department of Physics \& Astronomy, Michigan State University, East Lansing, Michigan, 48824, USA}

\author{Ann-Cecilie Larsen}
    \affiliation{Department of Physics, University of Oslo, Oslo, Norway}

\author{Rebecca Lewis}
    \affiliation{Facility for Rare Isotope Beams, Michigan State University, East Lansing, Michigan, 48824, USA}
	\affiliation{Department of Chemistry, Michigan State University, East Lansing, Michigan, 48824, USA}

\author{Stephanie Lyons}
    \affiliation{Pacific Northwest National Laboratory, Richland, Washington, 99354, USA}

\author{Alicia Palmisano-Kyle}
    \affiliation{Department of Nuclear Engineering, University of Tennessee, Knoxville, Tennessee, 37921, USA}

\author{Jorge Pereira}
    \affiliation{Facility for Rare Isotope Beams, Michigan State University, East Lansing, Michigan, 48824, USA}
    
\author{Andrea L. Richard}
    \affiliation{Physics and Astronomy Department, Ohio University, Athens, Ohio, 45701, USA}
    \affiliation{Facility for Rare Isotope Beams, Michigan State University, East Lansing, Michigan, 48824, USA}
    
\author{Debra Richman}
    \affiliation{Facility for Rare Isotope Beams, Michigan State University, East Lansing, Michigan, 48824, USA}
    
\author{Nicholas Scielzo}
    \affiliation{Lawrence Livermore National Laboratory, Livermore, California, 94551, USA}

\author{Anna Simon}
    \affiliation{Department of Physics and Astronomy, University of Notre Dame, Notre Dame, Indiana, 46556, USA}

\author{Mallory K. Smith}
    \affiliation{Facility for Rare Isotope Beams, Michigan State University, East Lansing, Michigan, 48824, USA}

\author{Chris Sullivan}
    \affiliation{Facility for Rare Isotope Beams, Michigan State University, East Lansing, Michigan, 48824, USA}
	\affiliation{Department of Physics \& Astronomy, Michigan State University, East Lansing, Michigan, 48824, USA}

\author{Adriana Sweet}
    \affiliation{Lawrence Livermore National Laboratory, Livermore, California, 94551, USA}

\author{Antonius Torode}
    \affiliation{Applied Research Laboratories, University of Texas, Austin, Texas, 78712, USA}
    \affiliation{Facility for Rare Isotope Beams, Michigan State University, East Lansing, Michigan, 48824, USA}
	\affiliation{Department of Physics \& Astronomy, Michigan State University, East Lansing, Michigan, 48824, USA}

\author{Remco Zegers}
    \affiliation{Facility for Rare Isotope Beams, Michigan State University, East Lansing, Michigan, 48824, USA}
	\affiliation{Department of Physics \& Astronomy, Michigan State University, East Lansing, Michigan, 48824, USA}

\date{\today}
\begin{abstract}

\end{abstract}
 
\maketitle

*\href{mailto:cdembski@nd.edu}{cdembski@nd.edu}
\newline
\textsuperscript{\textdagger}\href{mailto:spyrou@frib.msu.edu}{spyrou@frib.msu.edu}


\noindent

\section{Abstract}
The shape of the atomic nucleus is a property which underpins our understanding of nuclear systems, impacts the limits of nuclear existence, and enables probes of physics beyond the Standard Model. Nuclei can adopt a variety of shapes, including spheres, axially deformed spheroids, and pear shapes. 
In some regions of the nuclear chart where a spherical nucleus would naively be expected, deformed nuclear states can result from collective action of constituent protons and neutrons. In a small subset of nuclei both spherical and deformed nuclear states have been experimentally observed, a phenomenon termed shape coexistence. 
We present spectroscopic evidence for the coexistence of $J^{\pi}=1+$ spherical and deformed states in $^{70}$Co, separated by less than 275~keV. This close degeneracy of levels with the same $J^{\pi}$ and different shapes demonstrates an extreme example of shape coexistence resulting from the interplay of independent particle motion and collective behavior in highly unstable nuclear systems and identifies the Co isotopes as a transition point between deformed ground states observed in the Cr isotopes and spherical configurations observed in the closed-shell Ni isotopes.

\section{Introduction}
\label{sec:intro}
The nuclear shell model anchors our understanding of the atomic nucleus. It was developed 75 years ago by Mayer and Jensen \cite{may45,hax49} leading to the 1963 Nobel Prize in Physics and considers a single nucleon (either a neutron or a proton) moving in a mean field generated by the other nucleons.  The emergence of shell structure is a common occurrence in interacting quantum systems \cite{boh13,tar96,kni84} and provides a straightforward explanation for the experimentally observed “magic numbers” of 2, 8, 20, 28, 50, 82, and (for neutrons) 126. These numbers correspond to locations of large energy gaps between groups of single-particle states. Hence, completely filling the states up to one of these energy gaps imparts an extra stability to the atomic nucleus relative to its immediate neighbors, analogous to the extra chemical stability encountered in the noble gases when an atomic electron shell is filled. Nuclei located near these closed shells are predominantly spherical in shape; however, promoting just a few nucleons across the large energy gap can induce a change in the shape of the nucleus, giving rise to the phenomenon of ``shape coexistence''. The idea that the shape of the nucleus could change not only due to a different number of protons and neutrons but also from one state to another within the same nucleus was introduced almost 70 years ago for the case of $^{16}$O by Morinaga \cite{mor56}. Since then, shape coexistence has been observed experimentally in more nuclei \cite{gar22} and continues to provide a stringent test of modern nuclear structure models. Studies of the shape of the atomic nucleus have been instrumental in a diverse set of areas, including nuclear structure \cite{gaf13, tan19, ich19, rep21}, explorations of the limits of nuclear stability \cite{tsu20, smi23}, and even in physics beyond the Standard Model \cite{gaf13}.

In the original study of $^{16}$O\cite{mor56}, the coexisting states were the spherical $0^+$ ground state and a deformed $0^+$ excited state at an excitation energy of 6.05~MeV, resulting from the excitation of multiple particles across the large shell gap. Such a large energy separation is nominally expected from the nature of the excitation, but other observations have revealed a more varied picture. Generally, the relative energy between the spherical and deformed states is a result of a delicate balance between the energy cost of promoting particles across the shell gap and the energy gained from residual nucleon-nucleon interaction; The energy gained can more than offset the cost of the excitation and lead to the shape coexisting state dropping near, or below, the expected ground state. Therefore, shape-coexisting states provide a unique opportunity to study particle interactions within the same nucleus but under different conditions (different shapes). 

Such shape coexisting states are often found in transitionary regions of the nuclear chart, between mostly spherical nuclei found near closed shells and
heavily deformed open shell nuclei. Here we investigate
nuclei in the $A=70$ region near the doubly magic $^{78}$Ni. 
Prior studies have suggested the prevalence of shape evolution in this region \cite{gar22}, but thorough experimental study has been challenging due to the severe neutron-richness of the nuclei involved, and many prior investigations have focused on even-even isotopes \cite{Crider2018,Prokop2015}. It is thought that these deformations are driven by a phenomenon known as Type II shell evolution, in which the traditional nuclear shell picture is restructured due to multiparticle-multihole excitations across shell gaps \cite{Otsuka2016,Mor17}. Shape effects in odd-odd nuclei, like $^{70}$Co, have been suggested as strong tests of the Type II shell evolution mechanism \cite{Urban2017}, and their study is further necessary to understand broader scale transitionary effects that characterize nuclear structure in this region, like the possible merging of the N=40 and proposed N=50 islands of inversion \cite{Nowacki2016}.

\begin{figure}[t!]
\includegraphics[width=0.95\columnwidth]{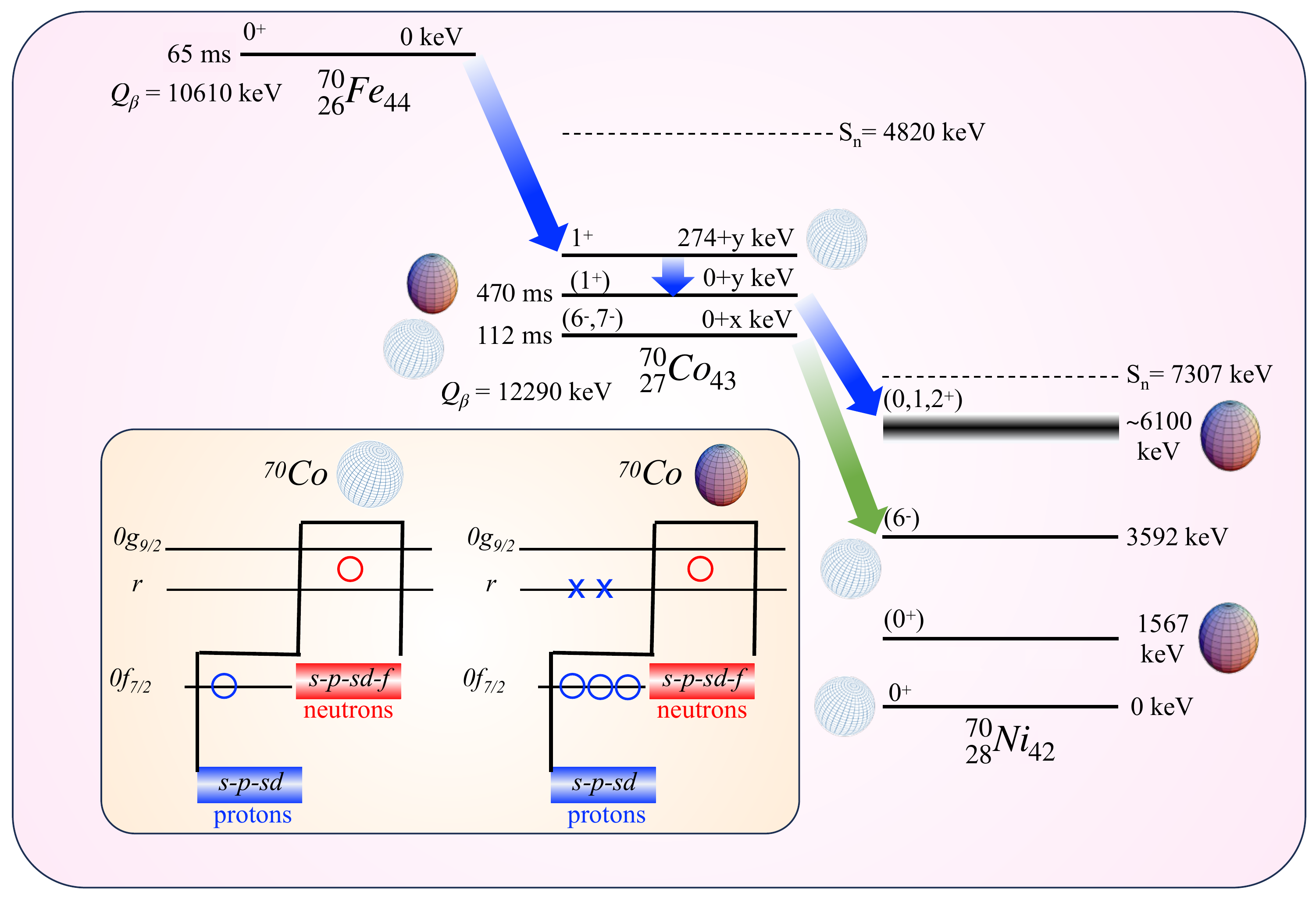}
\caption{\textbf{Dominant decay paths for $^{70}$Fe and $^{70}$Co.} The most prominent transitions observed in the decay chain under study in the present work, which pass through the low-spin isomer of $^{70}$Co, are shown as blue arrows. The green arrow shows the dominant transition from the high-spin isomeric state, observed in Spyrou et al.~\cite{Spy16} The inset shows the configurations of the two relevant $J^{\pi}=1^+$ states in $^{70}$Co. The label $r$ represents the group of orbitals $  (0f_{5/2},1p_{3/2},1p_{1/2}) $. Open circles represent hole states within the orbitals; x's represent states occupied by a particle excited across a shell gap. Symbols placed between orbitals indicate that configurations involving hole states in either orbital are possible.
States interpreted as having a weakly deformed (WD) character are labeled with a white spherical shape, while states interpreted as having a strongly deformed (SD) character are labeled with a  purple prolate shape (further details in text). All values are taken from the A=70 Nuclear Data Sheets \cite{Gurdal2016}, and the $J^{\pi}$ value of the long-lived $\beta$-decaying state in $^{70}$Co has been reassigned to $1^+$ in accordance with Morales et al.\cite{Mor17} and the Discussion within the Results section.}
\label{fig:decay_paths}
\end{figure}


Experimental investigation of the $^{70}$Fe$\rightarrow ^{70}$Co $\rightarrow ^{70}$Ni $\beta$-decay chain was possible using the technique of total absorption spectroscopy (TAS). The TAS technique was introduced almost 50 years ago \cite{har77} as a solution to the so-called ``Pandemonium Effect'', which refers to the erroneous assignment of $\beta$-decay feeding intensity ($I_\beta$) to lower excitation energies due to limitations in the detection efficiency and the presence of weak $\gamma$-ray transitions. TAS measurements avoid this issue by making use of large volume, high efficiency $\gamma$-ray calorimeters which allow for the accurate determination of the populated excitation energy, and hence accurate extraction of $I_\beta$. TAS detectors have been used extensively for $I_\beta$ measurements, and in particular for extracting nuclear shapes. For some nuclei, changes in deformation result in a marked difference in the Gamow-Teller strength distribution\cite{Sarriguren2001}, \textit{B}(GT), which is derived experimentally from $I_\beta$. Previous studies have compared TAS-measured and theoretical \textit{B}(GT) distributions to infer the deformations of  $\beta$-decaying states in $^{76}$Sr\cite{Nacher2004}, $^{78}$Sr\cite{Perez2013}, $^{78}$Rb\cite{Perez2013},  $^{100,101,102}$Zr\cite{Guadilla2019,Dombos2021}, and $^{109}$Tc\cite{Dombos2021}. This method is optimal for the study of $^{70}$Co, which has been found to have two $\beta$-decaying states \cite{Mueller2000}: one with high-spin ($J^{\pi}=6^-,7^-$) and a half life of $\approx 65$ ms \cite{Gurdal2016}, and a low-spin state with a half life of 
$\approx 470$ ms \cite{Gurdal2016}.
TAS measurement of the high-spin state was published by our group previously \cite{Spy16}, in excellent agreement with shell-model calculations showing its weakly deformed character. However, the low-spin state had not been studied in the same manner until now.

In the present work, a $^{70}$Fe beam was implanted at the center of a TAS detector called the Summing NaI(Tl) (SuN) detector \cite{Sim13}. The $\beta$-decay of the $J^{\pi}=0+$ ground state was used to preferentially populate the low-spin $\beta$-decaying state of $^{70}$Co and fully detail its decay pattern with minimal limitations from the Pandemonium Effect. 
The observed distributions, summarized in Fig.~\ref{fig:decay_paths}, are interpreted quantitatively with the assistance of state-of-the-art shell model calculations and strongly support an assignment of $J^{\pi}=1^+$ and a strong deformation for the low-spin $\beta$-decaying state, which has a wavefunction dominated by multiparticle-multihole excitations  across the Z=28 shell gap. Our results extend upon indications presented in Morales et al. \cite{Mor17} and identify an extreme case of shape coexistence between the deformed $J^{\pi}=1^+$ $\beta$-decaying state and near-spherical $J^{\pi}=1^+$ excited state $\sim$275~keV higher in excitation energy.

\section{Results}
\label{sec:results}

\subsection*{Experimental Results and Shell Model Calculations}
\begin{figure}[t!]
\includegraphics[width=0.95\columnwidth]{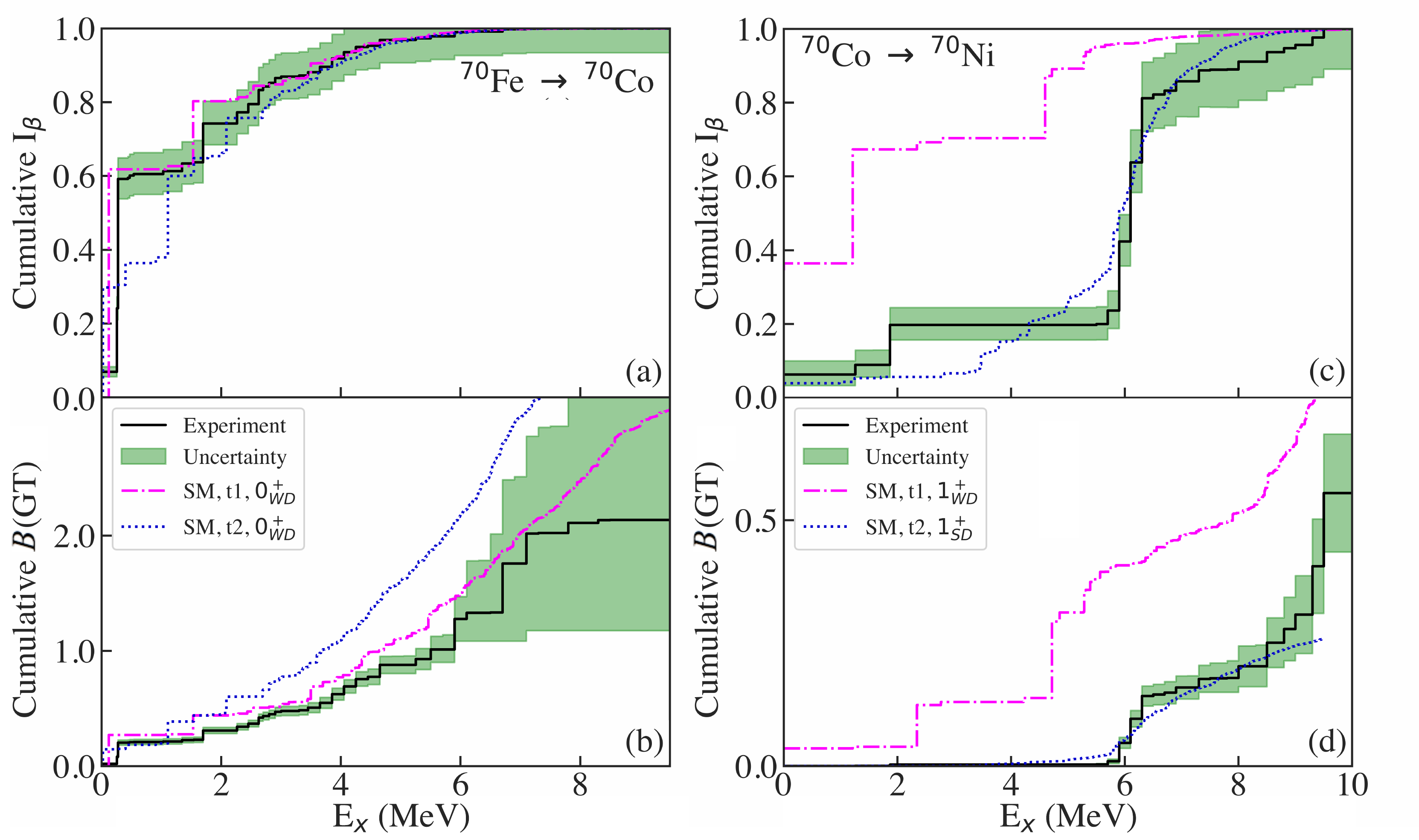}
\caption{\textbf{Cumulative $\beta$-Intensity and Gamow-Teller strength as a function of excitation energy for $^{70}$Fe and $^{70}$Co.} $^{70}$Fe decays are shown in panels (a) and (b); $^{70}$Co decays are shown in panels (c) and (d).  Experimental measurements are shown as a black line with a green band representing the uncertainty. For both nuclei, shell model calculations performed using the $t1$ truncation are shown as magenta dash-dot lines, while $t2$ truncation calculations are shown as a blue dotted line. Calculations are also labeled  by the shape of the $\beta$-decaying state.} 
\label{fig:betadecay}
\end{figure}

Experimentally measured and theoretically calculated $I_\beta$ and the extracted \textit{B}(GT) distributions for the two decays are shown in Fig.~\ref{fig:betadecay}. It can be observed that the $^{70}$Fe $\beta$-decay predominantly populates a level in $^{70}$Co that decays via emission of a 274~keV $\gamma$-ray. Following this  $\gamma$-deexcitation, the low-spin isomeric state in $^{70}$Co $\beta$-decays in a surprisingly selective pattern, predominantly feeding a highly localized set of levels with excitation energy of roughly 6 MeV in $^{70}$Ni. To connect these unique decay patterns to a microscopic interpretation of the nuclear shape, we performed shell model calculations of the same $\beta$-decay chain.

\begin{figure}
\includegraphics[width=0.6\columnwidth]{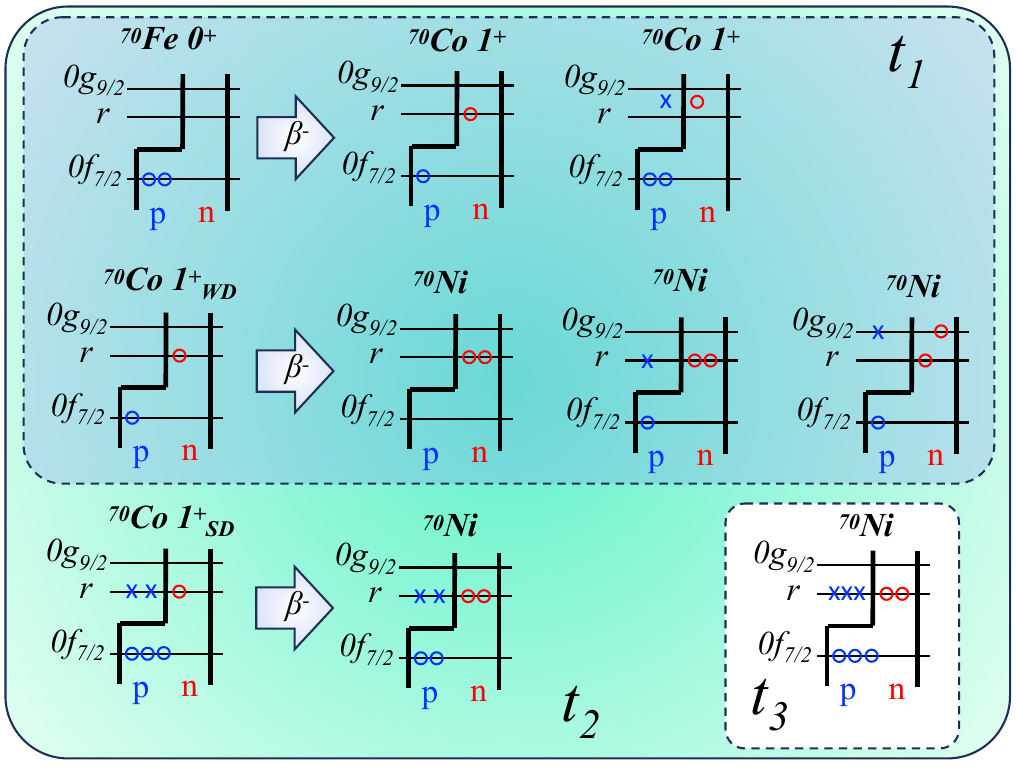}
\caption{\textbf{Basic possible configurations for the $\beta$-decays of $^{70}$Fe and $^{70}$Co.}
The top two rows (blue box) show the configurations related to the $  t1  $ truncation. Particle excitations are marked with x, while circles represent holes. Symbols placed between orbitals indicate that configurations involving particle/hole states in either orbital are possible. Orbitals are filled up to hole states. The last row shows the additional configurations allowed in the $  t2  $ truncation (green box). The $t3$ truncation, which is not included in the present calculations, would have the configuration shown in the white box.}
\label{fig:config}
\end{figure}

Gamow-Teller (GT) $\beta$-decay calculations were carried out with the NuShellX Code\cite{nushellx}. They employ a model space called $fpg9tn$ that consists of the five proton orbitals $  (0f_{7/2}, r, 0g_{9/2})  $ and four neutron orbitals $  (r,0g_{9/2})  $, with $r$ representing the $  (0f_{5/2},1p_{3/2},1p_{1/2})$ orbital group. Due to the computational complexity of the calculations, truncations to the model space were necessary; The label “$tn$” indicates that at most $n$ protons were allowed to be excited across the Z=28 shell gap (i.e.~from $0f_{7/2}$ to $(r,0g_{9/2}$)).  Fig.~\ref{fig:config} shows some simple proton-neutron configurations allowed under the model space truncations used. 
Calculated $T=1$ single-particle energies and two-body matrix elements for the $(r, 0g_{9/2})$ neutron orbitals were replaced with adjusted calculations using the $jj44a$ Hamiltonian to improve agreement of the spectra with the Ni isotopes and reproduce relative binding energies of $^{70}$Ni. More details about the Hamiltonians used can be found in the "Shell Model Hamiltonians" subsection of the Methods. 

\subsection*{Discussion}
\begin{figure}
\includegraphics[width=0.9\columnwidth]{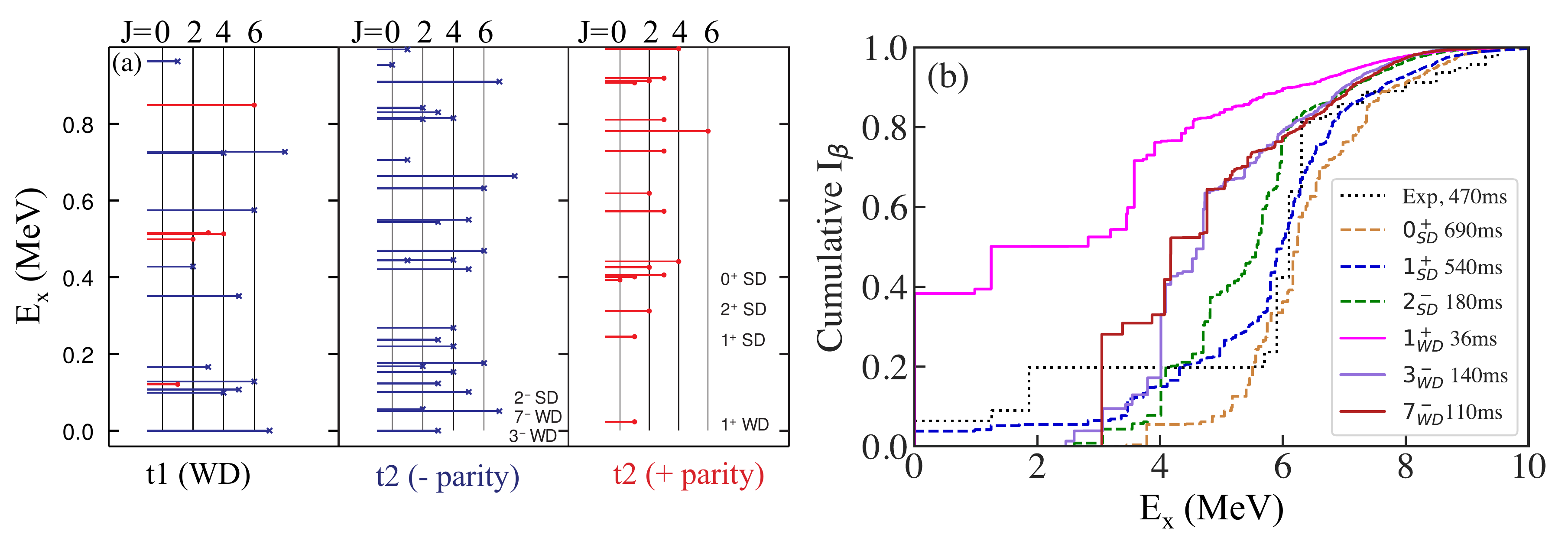}
\caption{\textbf{Calculated energy levels below one MeV in $^{70}$Co and associated $\beta$-decay patterns.} 
Panel (a) shows the calculated states. The length of the lines is
proportional to the J value and the color represents the parity (blue for negative, red for positive). States discussed in the text are labeled by their properties. Panel (b) shows The cumulative $\beta$-decay intensities and T$_{1/2}$ for the $\beta$-decay of $^{70}$Co to $^{70}$Ni based on possible band-head J$^\pi$ values in $^{70}$Co.  Decays from strongly deformed states are shown as dashed lines, while decays from weakly deformed states are shown as solid lines. The experimental intensity is shown as a dotted black line for comparison.}
\label{fig:sm}
\end{figure}

The Monte Carlo Shell Model calculations presented in Morales et al. \cite{Mor17} feature two major types of states: a group of low-spin, strongly deformed (SD) states at low energies, and a group of weakly deformed (WD) $J^{\pi} = (1^+,6^-,7^-)$  states at around 900 keV. These WD states are analogous to those we obtain using the $t1$ truncation. While this $t1$ truncation is able to reproduce many features of the $\beta$-decays observed in this mass region, including those of $^{70}$Fe (magenta dot-dash line in Fig.~\ref{fig:betadecay}a) and the high-spin $\beta$-decaying state of  $^{70}$Co\cite{Spy16}, it is not sufficient to accurately describe the observed $I_{\beta}$ (Fig.~\ref{fig:betadecay}c, magenta dot-dash line) or half-life (Fig.~\ref{fig:sm}b) of the $^{70}$Co low-spin $\beta$-decaying state of interest here. For this reason, the calculations were extended to the $t2$ truncation.

The $t2$ calculations predict 44 levels with energies less than 1 MeV, shown in Fig.~\ref{fig:sm}a, separated by positive (red) and negative (blue) parity. Many of these are rather pure SD states connected by strong B(E2) values, but a variety of low-lying WD states are also present. It is not clear \textit{a priori} which of these is the relevant $\beta$-decaying state. To elucidate this, $I_{\beta}$ and the $\beta$-decay half-lives (T$_{1/2}$) were calculated for candidates and compared with experiment, shown in Fig.~\ref{fig:sm}b. The decay from the WD $J^{\pi}=1^+$ state is similar to that calculated in the $t1$ truncation. The $J^{\pi}=7^-$ WD state corresponds to the short-lived, near-spherical $\beta$-decaying state discussed in Spyrou et al.\cite{Spy16} The lowest energy state in the calculations is a WD $J^{\pi}=3^-$ state, but both its decay pattern and half-life are inconsistent with experiment.

Decays from the SD states match the characteristics of the experimental distribution much more closely, exhibiting strong feeding to a range of states located between $5-7$ MeV. We interpret the Gamow-Teller decay of the SD $J^{\pi}=1^+$ state in $^{70}$Co to be populating
a few neutron-hole proton-particle Nilsson states in $^{70}$Ni (described further in the "Nilsson Model Interpretations" subsection of the Methods). The observed distribution of states comes
from the mixing of these simple configurations with the high density of other types
of $J^{\pi}=1^+$ states in $^{70}$Ni that are present around 6 MeV in excitation energy. The observed TAS peak at 6 MeV (see the Experimental Description subsection of the Methods) supports this interpretation, as it is wider than what would be expected for a single level in this region \cite{Sim13}. The $J^{\pi}=1^+$ SD  state has the best half-life agreement, and is the only SD  state whose decay features direct ground state feeding, which is observed experimentally in Fig.~\ref{fig:betadecay}c at an excitation energy of zero. This designation is also preferable because of the nonobservation of feeding to the lowest SD $J^{\pi}=0^+$ state in $^{70}$Ni. Experimentally, this state is located at 1.57 MeV\cite{Chiara2015} and is fed negligibly. In our SD $J^{\pi}=1^+$ calculations, the equivalent state appears at 3.2 MeV and is fed with only $0.8\%$ of the total intensity. Thus, our evidence is consistent with an assignment of $J^{\pi}=1^+$ and a strongly deformed nature for the low-spin, long-lived $\beta$-decaying state as suggested by Morales et al.\cite{Mor17}.

The two low-lying SD $J^{\pi}=1^+$ states that appear in our calculations also have wave function occupancies similar to those shown in Fig. 4 of Morales et al. \cite{Mor17} and large 2 particle-3 hole components. These states can be interpreted as members of deformed rotational bands with angular momentum projections of $K^{\pi}=(0^+,1^+)$. In these calculations, the $K^{\pi}=1^+$ band appears lower than the $K^{\pi}=0^+$ band. Adjusting the size of the Z=28 shell gap shifts the band energies relative to the predicted WD states, but they always appear close together and in the same order. A more detailed microscopic interpretation of these states using the Nilsson Model can be found in the "Nilsson Model Interpretations" subsection of the Methods.  

The WD $J^{\pi}=1^+$ state, appearing 220 keV below the SD $J^{\pi}=1^+$ state in our calculations, is also of significant importance. The experimental $^{70}$Fe $\beta$-decay (Fig.~\ref{fig:betadecay}a) prominently feeds a $J^{\pi}=1^+$ state that deexcites via a 274 keV $\gamma$-ray. This state is understood to be of a WD character - hence why it is populated so strongly in the $^{70}$Fe $\beta$-decay, while the SD $\beta$-decaying state is not despite having the same $J^{\pi}$. The reason that the WD state appears below the SD state in the calculations is due to a shift in the energies of SD states resulting from the model space truncation used. This is illustrated in the case of a “mirror” N=27 isotone, $^{51}$Cr, which is discussed in detail in "The N=27 'mirror' case" subsection of the Methods. This study is relevant since the $^{51}$Cr calculations can be performed in the full model space, providing an estimate of the limitations of the $tn$ truncations used for $^{70}$Co. $^{51}$Cr demonstrates a similar WD $J^{\pi}=7/2^-$ state and SD $K^{\pi}=1/2^-$ band in its lowest lying levels. Calculations that use the $t2$ neutron truncation see the $K^{\pi}=1/2^-$ SD band energies placed roughly 600 keV too high relative to the full space calculations. Correcting for this effect, the SD $\beta$-decaying state in $^{70}$Co would fall a few hundred keV below the strongly $\beta$-fed WD $J^{\pi}=1^+$ state, in excellent agreement with present experimental observations.

\section{Conclusions}
 We have presented spectroscopic evidence indicating that the low-spin $\beta$-decaying state in $^{70}$Co has $J^{\pi}=$ $1^+$ and a strongly deformed character. These results are consistent with and extend upon previous literature results, and identify one of the closest spacings between nuclear levels of the same $J^{\pi}$ and different shapes yet observed. Such  effects provide stringent tests for models at the limits of nuclear structure, and could exert influence on complex astrophysical nucleosynthesis pathways far from stability. Investigation of nearby nuclei, including those near to the N=50 shell closure, could reveal similar effects and is necessary for a complete understanding of the structural evolution in this region.
 In the case of $^{70}$Co, future experiments, such as measurement of the lifetime of the 274 keV level and/or the angular distribution of its deexcitation to probe the multipolarity of the transition would further confirm the present interpretation. Additional theoretical calculations for the $\beta$-decays that extend to a $  t3  $ truncation and include the $  1d_{5/2}  $ orbital also remain to be carried out.

\section*{References}
\nocite{apsrev41Control}
\bibliographystyle{apsrev4-1}
\bibliography{references}

\vspace{1cm}
\section{Methods}
\input{methods}

\section*{Methods References}
\nociteM{apsrev41Control}
\bibliographystyleM{apsrev4-1}
\bibliographyM{references}

\begin{acknowledgments}

This research used resources of the National Superconducting Cyclotron Laboratory (NSCL), which was an NSF supported User Facility. The work was supported by the National Science Foundation under grants
PHY 1913554,  
PHY 2209429, 
PHY 1565546, 
PHY 1848177 (CAREER) 
PHY 2110365, 
PHY 1613188, 
PHY 2310059. 
 This material is based upon work supported by the Department of Energy/National Nuclear Security Administration through the Nuclear Science and Security Consortium under Award Nos. DE-NA0003180 and DE-NA0003996, and Stewardship Science Academic Alliance under DE-NA0003906. This material is based upon work supported by the U.S. Department of Energy, Office of Science, Office of Nuclear Physics, under Contract Nos. DE-AC02-06CH11357, DE-SC0023633, DE-SC0020451, and DE-AC52-07NA27344.
 A.C.L. gratefully acknowledges funding from ERC-STG-2014 under Grant Agreement No. 637686, the Research Council of Norway, Project No. 316116, and support from the Norwegian Nuclear Research Centre, Project No. 341985.

\end{acknowledgments}

\section*{Data Availability}
The dataset presented in this manuscript will be made available upon request to the corresponding author(s).
\section*{Code Availability}
The code used for analysis will be made available upon request to the corresponding author(s).

\section*{Author Contributions}
C. Dembski performed the analysis of the experimental data, the interpretation of the results and the writing of the manuscript.\\
A. Spyrou worked on the proposal, preparation and execution of the experiment, the data analysis, the interpretation of the results and writing of the manuscript. \\
B.A. Brown performed the theoretical calculations and worked on the interpretation of the results and writing of the manuscript.\\
S.N. Liddick worked on the proposal, preparation and execution of the experiment, the interpretation of the results and writing of the manuscript.\\
H. C. Berg worked on the simulations of the $\gamma$-cascades and corresponding detector response, and helped with the manuscript preparation.\\
E.C. Good worked on the $\beta$-ion correlations, and preparation of the manuscript.\\
C. Harris, R. Lewis, S. Lyons, A. Palmisano-Kyle, A. Richard, D. Richman, and M.K. Smith prepared the experimental setup, installed and calibrated detectors, participated in the experiment execution, and the preparation of the manuscript. \\
D.L. Bleuel, K. Childers, B.P. Crider, A.C. Dombos, A.C. Larsen, J. Pereira, N. Scielzo, A. Simon, C. Sullivan, A. Sweet, A. Torode and R. Zegers participated in the experiment and online analysis and helped with the manuscript preparation.   \\

\section*{Competing Interests}
The authors declare no competing interests.

\end{document}

%% file: methods.tex
\label{methods}
\subsection{Experimental Description}
\label{sec:exp}
The experiment was conducted at the National Superconducting Cyclotron Laboratory at Michigan State University. The experimental setup was detailed in previous publications \protect\citeM{Lyons2019,Dom19} and will be summarized here.
A $^{86}$Kr primary beam at 140 MeV/u was impinged on a $^{9}$Be target, and the isotopes of interest were separated using the A1900 fragment separator \protect\citeM{Mor03}. Individual isotopes were identified using energy loss and time-of-flight information recorded by a plastic scintillator in the focal plane of the A1900, coupled with two silicon PIN detectors near the end-station. Following identification, the beam was implanted into a double-sided silicon strip detector (DSSD), which was used to detect the location and time for both the implanted ions and the electrons produced by their $\beta$-decays. $\gamma$-rays emitted following these $\beta$-decays were detected using the SuN detector \cite{Sim13}. SuN is a cylindrical scintillator detector divided into 8 optically isolated NaI(Tl) segments.  The segment signals can be evaluated individually (sum-of-segments), providing a measure of the $\gamma$-rays emitted during the deexcitation of a populated level, or summed over all 8 segments to provide a total absorption spectrum that is sensitive to the excitation energy of the nucleus. A third spectrum called "multiplicity" records the number of segments that register a signal in each event and is an indication of the number of $\gamma$-rays emitted from a deexcitation.

\begin{figure}
\includegraphics[width=0.7\columnwidth]{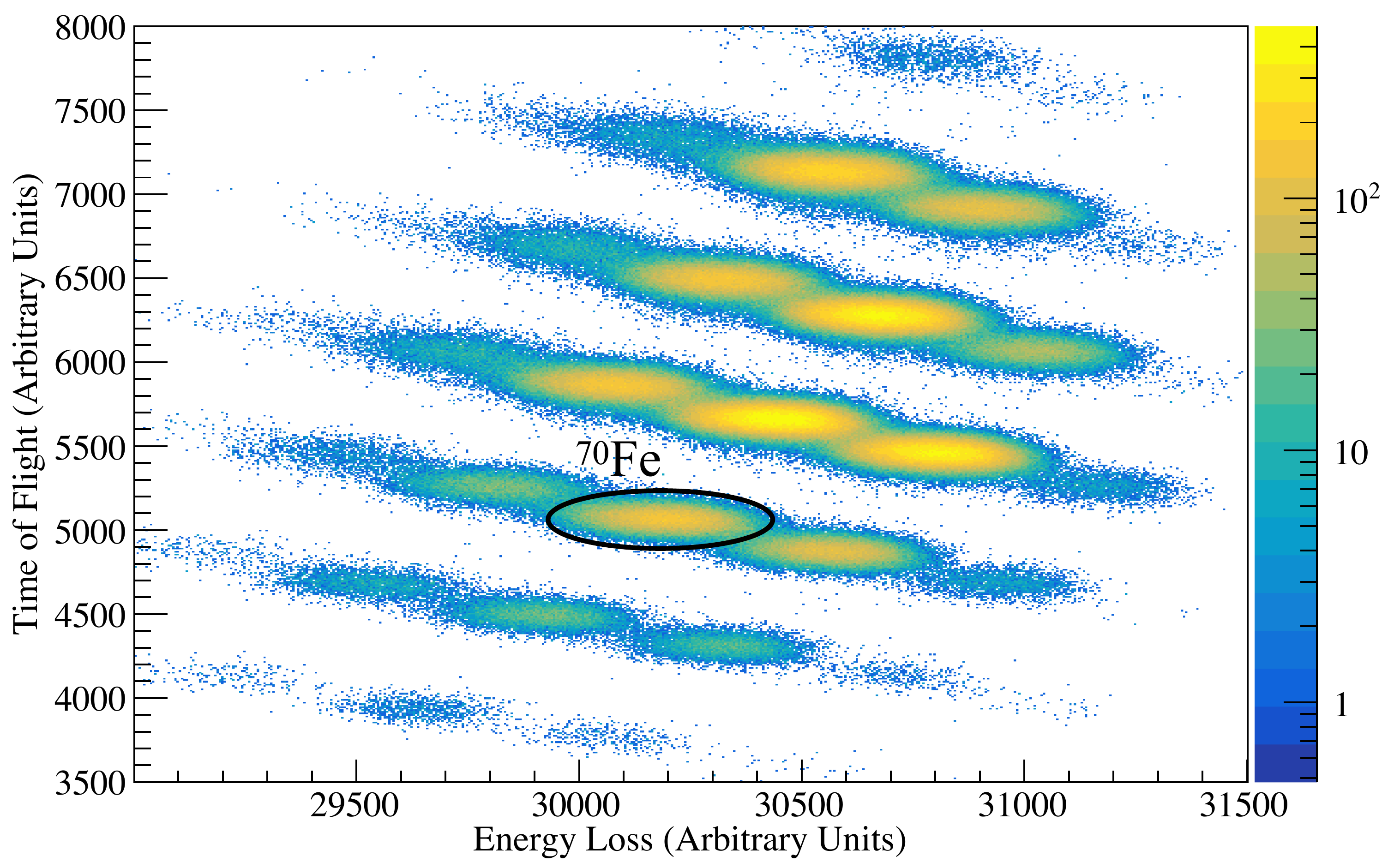}
\caption{\textbf{Particle identification plot showing time of flight and energy loss measurements.} The gate used to isolate A=70 fragments is also shown as a black oval. 
}
\label{fig:PID}
\end{figure}

Events corresponding to $^{70}$Fe ions were isolated by gating on the appropriate ranges for time-of-flight and energy loss measured by the silicon PIN detectors. The particle identification plot and gate are shown in Fig.~\ref{fig:PID}. The DSSD allowed for both spatial and temporal correlation of implantations with subsequent $\beta$-decays. Decays detected within 150 ms of implantation were assigned to $^{70}$Fe. Decays of the $^{70}$Co child nucleus were isolated using the same particle ID gate, but a timing window of 400-1000 ms following the $^{70}$Fe ion detection. Not every $\beta$ detection within the correlation requirements corresponds to a real decay from a particular ion; therefore, the correlation process was repeated with $\beta$-detections that precede the ion implantation to produce a random correlation background. This background was subtracted from the raw ion-$\beta$ correlated spectra produced for $^{70}$Fe decays. For $^{70}$Co decays, which have lower statistics, the background was included as a component of the fitting procedure outlined below and detailed thoroughly in Dombos et al. \protect\citeM{Dom16}

\begin{figure}[t!]
\includegraphics[width=0.95\columnwidth]{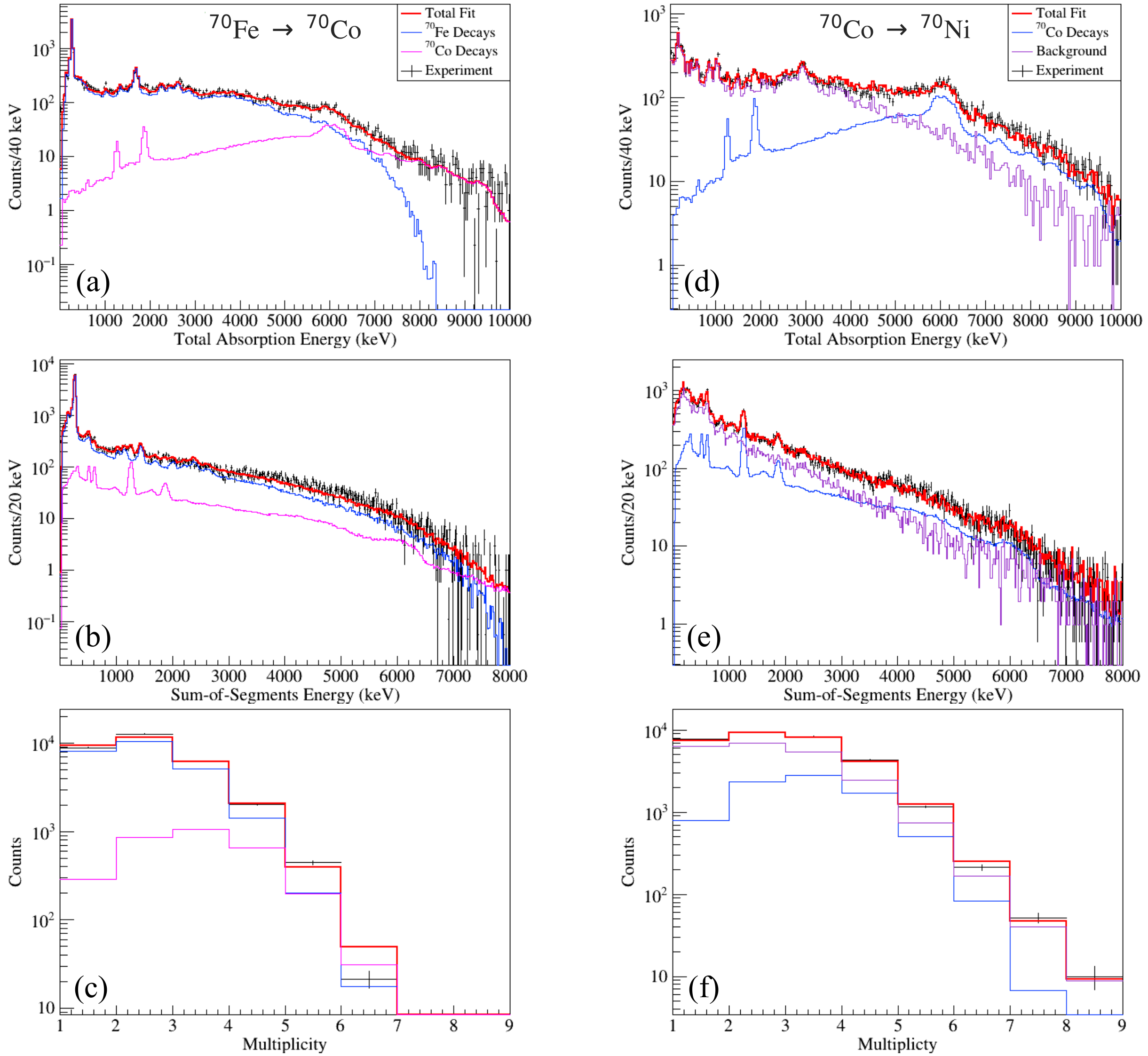}
\caption{\textbf{Total absorption, sum-of-segments, and multiplicity fits for $^{70}$Fe and $^{70}$Co decays.} 
For $^{70}$Fe, experimental spectra for $^{70}$Co child decays was included as a component in the fit and is shown separately (magenta, panels (a), (b), and (c)). For $^{70}$Co, the random background was included as a component of the $\chi^2$ minimization, instead of being subtracted prior to the fit, due to low statistics. It is shown in purple in panels (d), (e), and (f).}
\label{fig:fits}
\end{figure}

To extract the $\beta$-decay feeding intensity ($I_{\beta}$) information, a $\chi^2$ minimization method that simultaneously fits the total absorption, sum-of-segments and multiplicity spectra was employed. Simulated decays from each discrete level in the the child nucleus were created using a GEANT4\protect\citeM{Agostinelli2003} simulation of SuN \cite{Sim13}. For $^{70}$Co, the discrete level spectrum presented in Morales et al. \cite{Mor17} was used, and for $^{70}$Ni the levels were taken from the National Nuclear Data Center's A=70 Nuclear Data Sheets \protect\citeM{Gurdal2016}. For energies between the highest lying discrete level and Q$_{\beta}$, decays from pseudolevels spaced roughly equally to SuN's energy resolution were simulated using the software RAINIER \protect\citeM{Kirsch2018}. RAINIER uses a user-defined nuclear level density (NLD) to create an artificial level scheme and follows the deexcitation of the nucleus using a Monte Carlo method with a $\gamma$-ray strength function (gSF). The extracted $I_{\beta}$ is not very sensitive to small variations of the NLD or gSF. For the present work, the constant temperature NLD model \protect\citeM{Egi09} and the gSF extracted for $^{70}$Ni in Liddick et al.\protect\citeM{Lid16} and Larsen et al. \protect\citeM{Lar18} were used. It was assumed that simulated pseudo-levels were populated by allowed decays, therefore they were assigned $J^{\pi}=1^+$ in the decay $^{70}$Fe $\rightarrow$ $^{70}$Co and $J^{\pi}=0^+,1^+,2^+$ in the decay $^{70}$Co $\rightarrow$ $^{70}$Ni, using our assignment of $J^{\pi}=$ $1^+$ for the $\beta$-decaying state of $^{70}$Co.

For each discrete level and each pseudolevel, GEANT4 was used to simulate total absorption, sum-of-segments, and multiplicity spectra. These spectra were used as components in a $\chi^2$ fitting procedure that simultaneously fit the experimentally measured total absorption, sum-of-segments, and multiplicity spectra. The relative weight of each component determined the relative $I_{\beta}$ at the given excitation energy. Experimental fits for both nuclei are shown in Fig.~\ref{fig:fits}.

To account for unassigned $J^{\pi}$ values in the discrete level schemes of these nuclei, a set of 10 RAINIER simulations was done for each nucleus, where any unassigned level's spin was randomly varied within a $\Delta J$=2 window of existing $J^{\pi}$ estimates. For $^{70}$Fe, these estimates were taken from shell model calculations presented in Morales et al.\cite{Mor17} For $^{70}$Ni, spin ranges for a given level were estimated based on the production mechanisms in the National Nuclear Data Center's A=70 Nuclear Data Sheets\cite{Gurdal2016}. The results from the spin variation were included in the uncertainty estimate of $I_{\beta}$, along with statistical uncertainties and as a 10\% efficiency uncertainty\cite{Sim13}. 

The extracted I$_{\beta}$ distributions were used to calculate the associated Gamow-Teller strengths using 
\begin{equation}
    B(\text{GT})(E) = K(\frac{g_v}{g_a})^2\frac{I_{\beta}(E)}{f(Q_\beta-E)T_{\frac{1}{2}}}
\end{equation}
where $K$ = 6143.6(17) s \protect\citeM{Hardy2009}, $\frac{g_a}{g_v}$ = -1.2695(29) \protect\citeM{Yao2006}, $f(Q_{\beta}-E)$ is the Fermi Integral, and $T_\frac{1}{2}$ is the half-life.

\subsection{Shell Model Hamiltonians}
\label{sec:hamiltonians}
The starting Hamiltonian was obtained using the many-body perturbation theory code of Hjorth-Jensen et al.\protect\citeM{jensen}. We use the N$^3$LO two-nucleon interaction, renormalized for short-ranged correlations with the $  v_{lowk}  $ method with a cutoff of $  \Lambda =2.2  $ MeV \protect\citeM{Bogner2003}.
Two-body matrix elements were obtained for a basis of states that contained up to 4$\hbar\omega$ excitations beyond a $^{68}$Ni closed shell, where the $  0f_{7/2}  $ orbital is filled for protons
and the $  (r)  $ orbitals are filled for neutrons.
The radial integrals were calculated with harmonic-oscillator
wavefunctions with $\hbar\omega$ = 9.16 MeV.
The renormalized Hamiltonian for the $  fpg9  $ model space
included up to second-order core-polarization corrections with folded diagrams. 
	To improve agreement with the spectra of the Ni isotopes, the T=1 two-body matrix elements and single-particle energies for the ($r,0g_{9/2}$) neutron orbitals were replaced with those obtained in Lisetskiy et al\protect\citeM{z28} (called the \textit{jj}44\textit{a} Hamiltonian in Mukhopadhyay et al.\protect\citeM{jj44a}). This Hamiltonian was designed for use in the $t0$ truncation (i.e. a Z=28 closed shell). For the $t1$ truncation, the particle-hole components of the wavefunctions shifted the energies of states associated with the ($r,0g_{9/2}$) neutron configurations relative to those obtained with the \textit{jj}44\textit{a} Hamiltonian. To reconcile this difference, the \textit{jj}44\textit{a} single-particle energies were adjusted and the two-body matrix elements were normalized by a factor of 0.80. The proton single-particle energies were adjusted to approximately reproduce the relative binding energies of $^{69}$Co, $^{70}$Ni, and $^{71}$Cu, and the low-lying states of $^{71}$Cu. 
	All \textit{B}(GT) calculations using this final Hamiltonian were carried out assuming a standard Quenching factor of $R=0.60$.  This is a typical reduction factor required for Gamow-Teller strengths observed in $\beta$-decays when compared to shell model calculations in the \textit{sd}\protect\citeM{rds} and \textit{fp}\protect\citeM{rfp}. A similar Quenching factor can be extracted from (p,n) reactions when compared to the $3(N-Z)$ Gamow-Teller Sum Rule.\protect\citeM{pn}

\subsection{Nilsson Model Interpretations}
\label{sec:nilsson}
There is small mixing between the WD $J^{\pi}$=$1^+$ and SD $J^{\pi}$=$1^+$ states in our calculations for $^{70}$Co. This can be interpreted in terms of the Nilsson Model\protect\citeM{Nilsson1955} and the potential energy surface (PES) shown in Fig. 3 of Morales et al.\cite{Mor17}. The WD $J^{\pi}$=$1^+$ state is dominated by the coupling of the spherical $\pi0f_{7/2}$ and $\nu0f_{5/2}$ configurations. However, as deformation increases, the proton $  [N,n_{z},\Lambda ]\mid \Omega _{p}\mid ^{\pi }=[3,0,3]7/2^{-}  $ and $  [3,2,1]1/2^{-}  $ Nilsson orbitals cross, resulting in a new PES minimum with a deformation of $  \beta  \approx +0.3  $. The proton 1 particle-1 hole states at the $t1$ level of truncation can describe small changes around $\beta_{min}$=0, but are not enough to describe these states with $\beta=+0.3$; for such large deformations the $t2$ truncation is necessary.

\begin{figure}
\includegraphics[width=0.6\columnwidth]{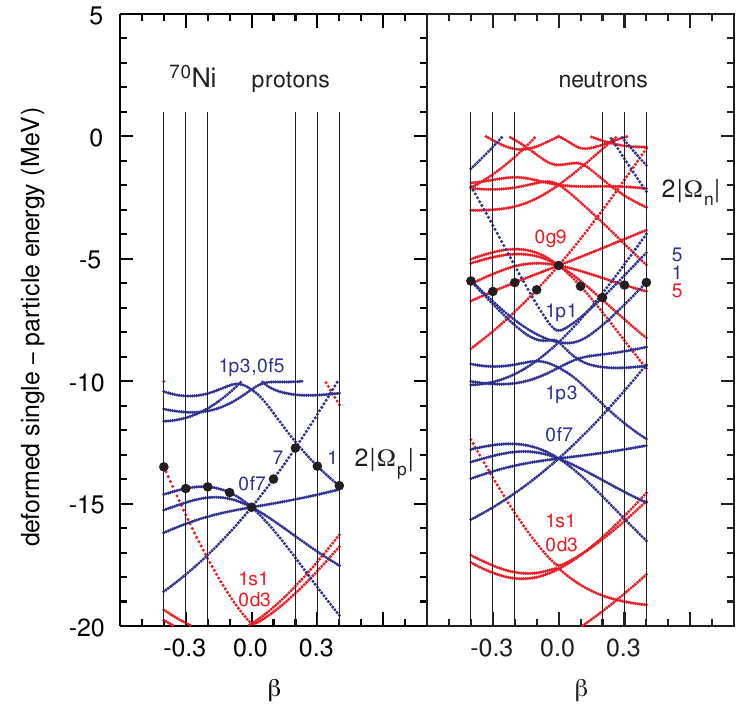}
\caption{\textbf{Nilsson diagram for $^{70}$Ni.} Calculations are performed with the
deformed Woods-Saxon code WSBETA \protect\citeM{dws}
with the potential choice ICHOIC=3.
Positive-parity states are shown in red and
negative-parity states are shown in blue.
The proton
and neutron 2$\mid\Omega\mid^{ \pi }$ values  that
are related to the deformed states in $^{69,70}$Co
discussed in the text are labeled.}
\label{fig:nilsson}
\end{figure}

The Nilsson Model is also useful in understanding the origins of the WD and SD  states in these nuclei. A proton Nilsson diagram for $^{70}$Ni is shown in Fig.~\ref{fig:nilsson}, with protons filling the $0f_{7/2}$ orbital at small deformation. As such, one might expect the ground state of neighboring $^{69}$Co to be well-described by one proton hole in the $  0f_{7/2}  $ orbital with $  J^{\pi}=7/2^{-}  $. As the deformation increases, the
 $  \mid \Omega _{p}\mid ^{\pi }=7/2^{-}  $ and $  \mid \Omega _{p}\mid ^{\pi }=1/2^{-}  $ Nilsson 
 orbitals cross,
 so that near $\beta$=$+$0.3,  the Nilsson state just below the Fermi surface has
 $  \mid \Omega _{p}\mid ^{\pi }=1/2^{-}  $, arising from a proton
 2 particle-2 hole structure in $^{70}$Ni. Removing a proton would now result in a deformed $  K^{\pi}=1/2^{-}  $ rotational band. In calculations of $^{69}$Co using the $t2$ truncation, shown in Fig.\ref{fig:BE2_cont}(b), one sees the expected $  J^{\pi}=7/2^{-}  $ ground state and $  K^{\pi}=1/2^{-}  $ rotational band, dominated by 2 particle-3 hole configurations (although due to staggering its lowest-lying state has $  J^{\pi}=3/2^{-}  $, and also may be a $\beta$-decaying isomer). Experimentally, low-lying states of $J^{\pi} = 7/2^-$\protect\citeM{Mue99} and $J^{\pi} = 1/2^-$\protect\citeM{lid15} have been observed that may be associated with these WD and SD shapes. Calculations for $^{69,71,73}$Co with the LNPS Hamiltonian in a model space similar to ours are discussed in Lokotko et al\protect\citeM{Lokotko2020}. Neutron orbitals exhibit the same pattern; In $^{43}$S, one observes an SD $  K^{\pi }=1/2-  $ band and WD $J^{\pi}=7/2^{-}$ state associated with the $  0f_{7/2}-r  $ neutron configurations \protect\citeM{Longfellow2020}.

 A clear pattern appears that can also describe the low-lying states observed in $^{70}$Co. With our $  t1  $ Hamiltonian the lowest negative-parity state in $^{70}$Co is 7$^{-}$,
 and the lowest positive-parity state is 1$^{ + }$.
 The dominant configurations of these states are
 $  (\pi 7/2^{-} \otimes \nu 9/2^{+})  $, and $  (\pi 7/2^{-} \otimes \nu 5/2^{-})  $, respectively.
Extending to the $t2$ truncation allows for stronger deformation, and thus the aforementioned crossing of the $  \mid \Omega _{p}\mid ^{\pi }=7/2^{-}  $ and $  \mid \Omega _{p}\mid ^{\pi }=1/2^{-}  $ orbitals. In this model space, SD positive-parity states in $^{70}$Co can be interpreted as the
 coupling of the $\mid\Omega_{p}\mid^{ \pi }$=1/2$^{-}$ and $\mid\Omega_{n}\mid^{ \pi 
 }$=(1/2,5/2)$^{-}$
 Nilsson states to make proton-neutron bands with $  K=(0,1,2,3)^{+}  $.
 SD negative-parity states in $^{70}$Co can be interpreted as the
 coupling of the of the $\mid\Omega_{p}\mid^{ \pi }$=1/2$^{-}$ and $\mid\Omega_{n}\mid^{ \pi 
 }$=5/2$^{ + }$
 Nilsson state to make proton-neutron bands with $  K=(2,3)^{-}  $.

Finally, the Nilsson model provides a reason for the suppression of the nominally allowed $\beta$-decay transition to the SD $J^{\pi}$=$0^+$ discrete state at 1.57 MeV $^{70}$Ni. In our calculations, this decay is dominated by the forbidden neutron $  [3,2,1]1/2^{-}  $
to proton $  [3,0,1]1/2^{-}  $ GT transition.

\subsection{The N=27 "mirror" case}
\label{sec:n27}

The connection between the model-space truncation and the
energies of the SD bands can be studied
in $^{51}$Cr, which falls analogously below the $N=28$ shell gap and can be fully calculated with no truncations to the model space. States starting with the 3/2$^{-}$ level at 0.75 MeV
and the 1/2$^{-}$ level at 0.78 MeV \protect\citeM{cr51,Heyde1983}
can be interpreted as the start of a  $  K^{\pi }=1/2^{-}  $
SD band based on the neutron $  [3,2,1] 1/2^{-}  $ Nilsson orbtial.
The levels of $^{51}$Cr can be described within the $  (0f_{7/2},r)  $ model space
with the well-established GPFX1A Hamiltonian
with effective charges of $  e_{p}=1.5  $ and $  e_{n}=0.5  $ \protect\citeM{gx1a}.
This collective band is formed from
exciting neutrons from $  0f_{7/2}  $ to the $  r  $ subset of orbitals.
The WD and SD properties of $^{51}$Cr can be studied with an E2-map, where
the levels are connected by lines whose width is proportional
to the electric quadrupole transition matrix element $|M(E2)|^2$, and the transition strength is given by B(E2) =  $|M(E2)|^2/(2J_i+1)$. The results of the full-space calculation
are shown in Fig. \ref{fig:BE2}(d). The energies and
B(E2) are in excellent agreement with experimental measurements\protect\citeM{cr51,Heyde1983}.
These full-space results are compared to those from
various levels of truncation $  tn  $ in other panels of Fig. \ref{fig:BE2},
where $  n  $ is the maximum number of neutrons allowed in the
$  r  $ subset of orbitals.
Weakly deformed (WD) states that are associated with
those that start with the 7/2$^{-}$ ground state
are present even at the $  t1  $
level of truncation. On the other hand, SD rotational bands do not appear until excitation of multiple neutrons across the shell gap is allowed in the $t2$ truncation.


\begin{figure}
\includegraphics[width=0.95\columnwidth]{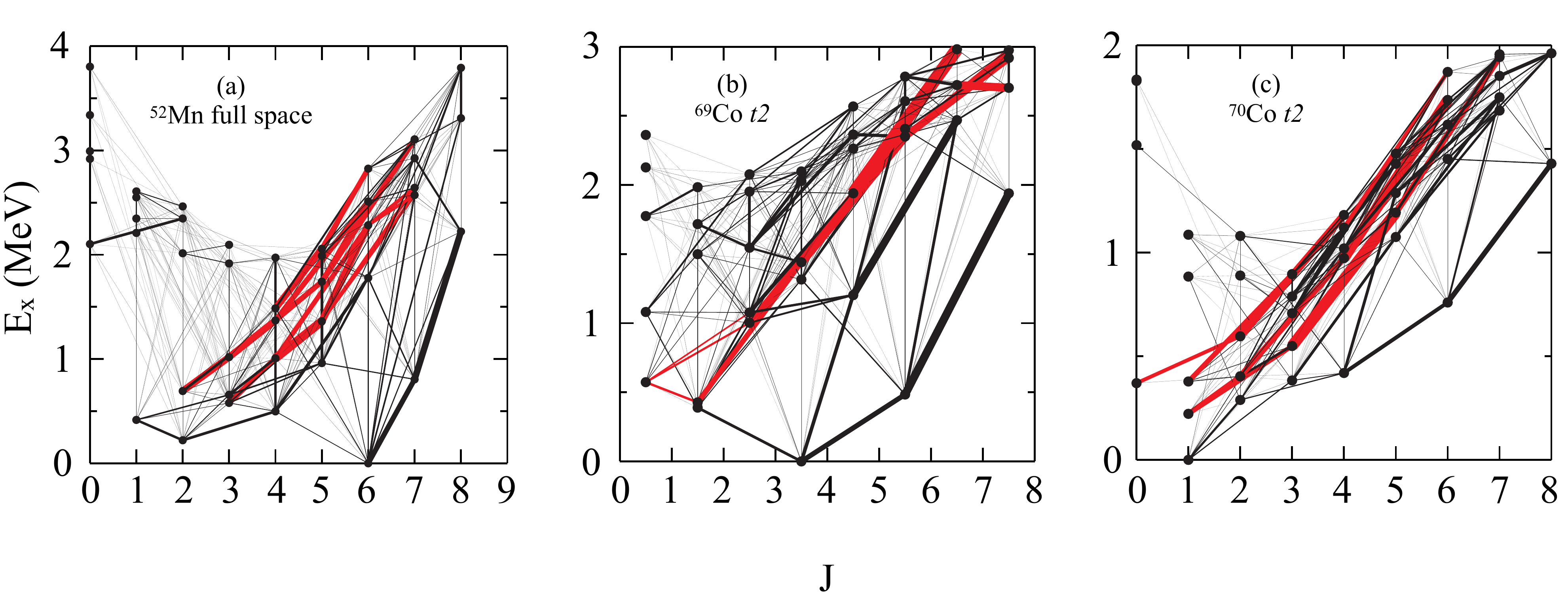}
\caption{\textbf{E2-maps for $^{52}$Mn, $^{69}$Co, and $^{70}$Co.} Panel (a) shows full model space calculations for $^{52}$Mn.  Panels (b) and (c) show calculations using the $t2$ truncation for $^{69}$Co and $^{70}$Co, respectively. The width of the lines connecting
the levels is proportional to the $|M(E2)|^2$. Strong transitions involved in the $K^{\pi}=2^+,3^+$ bands in (a), $K^{\pi}=1/2^+$ band in (b), and $K^{\pi}=0^+,1^+$ bands in (c) are shown in red.}
\label{fig:BE2_cont}
\end{figure}
Comparison between the $^{69,70}$Co neutron neighbors and $^{51}$Cr,$^{52}$Mn proton neighbors is useful in completing this picture; In odd-odd $^{52}$Mn there are SD states starting with $J^{\pi} = 2^+$ and $3^+$ shown in 
Fig. \ref{fig:BE2_cont}(a) that can be interpreted as the coupling of the neutron $[3,2,1]1/2^-$ and proton $[3,1,2]5/2^-$ Nilsson states. 
These coexisting sets of states $^{51}$Cr and $^{52}$Mn are analogous to those in 
$^{69}$Co (shown in Fig.\ref{fig:BE2_cont}(b))  and $^{70}$Co (shown in Fig.{\ref{fig:BE2_cont}(c)}), which are the ultimate focus of this paper.
The deformed band in $^{51}$Cr obtained with the $t2$ truncation is 600 keV higher than that obtained
with the full model space calculation. Given the similarities between the two pairs, we can expect the energy of the deformed bands in  $^{69,70}$Co
obtained with the $t2$ truncation to be too high in energy by a similar amount.
Taking this into account, the SD $J^{\pi}=1^+$ state would be lower in energy than the WD $J_{\pi}=1^+$ state,
in agreement with the present experimental observation.

\begin{figure}
\includegraphics[width=0.7\columnwidth]{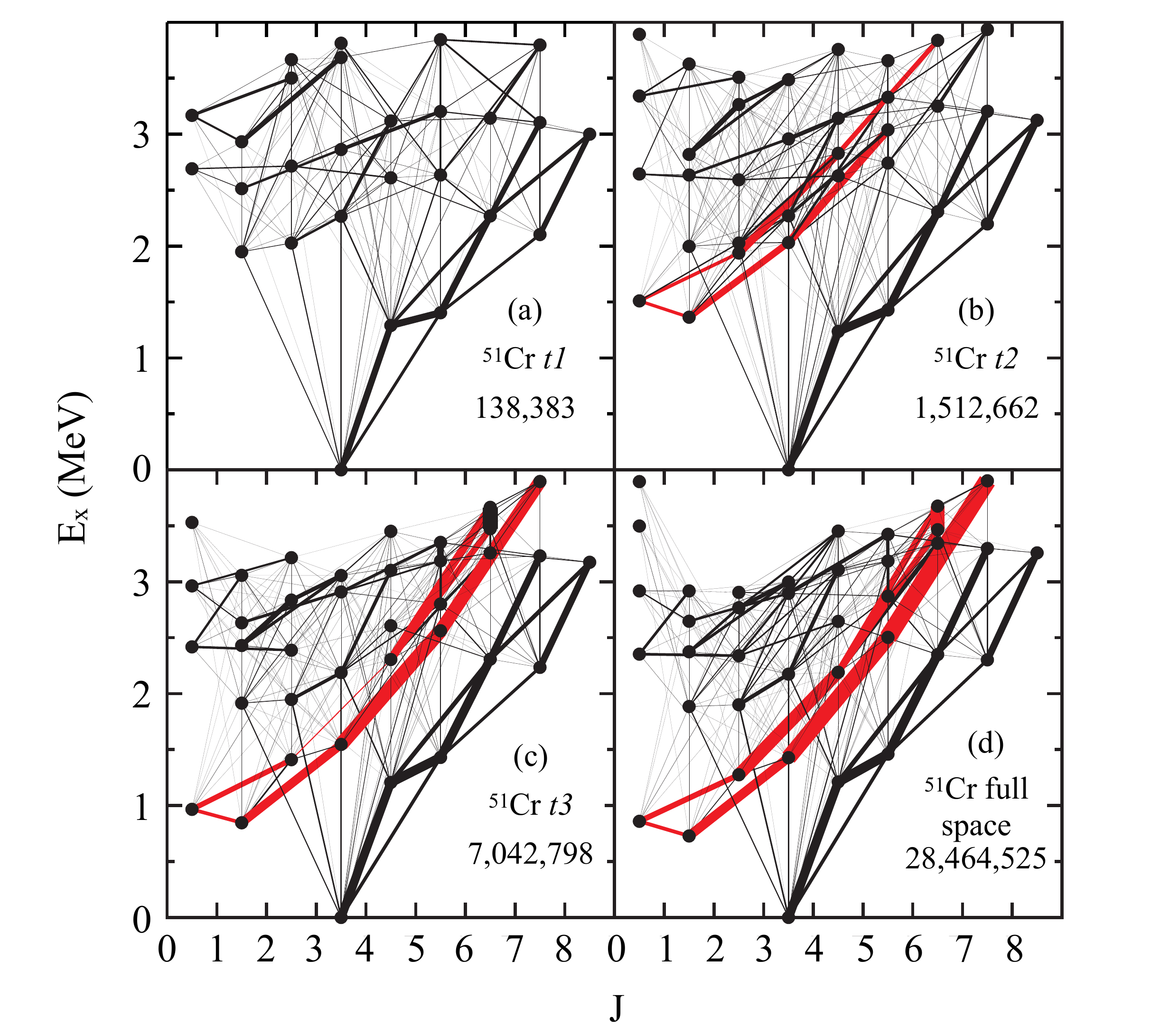}
\caption{\textbf{E2-map for $^{51}$Cr.} Panel (a) shows the E2-map calculated using the $t1$ truncation. Panels (b), (c), and (d) show the same, but utilizing the $t2$ truncation, $t3$ truncation, and full model space, respectively.
The width of the lines connecting the levels is proportional to the $|M(E2)|^2$, and the strong
transitions involved in the $  K^{\pi }=1/2^{-}  $ band
are shown in red. The number in each panel is the $  m  $-scheme
dimension for $M=1/2$.}
\label{fig:BE2}
\end{figure}

%% file: main.bbl
\begin{thebibliography}{28}%
\makeatletter
\providecommand \@ifxundefined [1]{%
 \@ifx{#1\undefined}
}%
\providecommand \@ifnum [1]{%
 \ifnum #1\expandafter \@firstoftwo
 \else \expandafter \@secondoftwo
 \fi
}%
\providecommand \@ifx [1]{%
 \ifx #1\expandafter \@firstoftwo
 \else \expandafter \@secondoftwo
 \fi
}%
\providecommand \natexlab [1]{#1}%
\providecommand \enquote  [1]{``#1''}%
\providecommand \bibnamefont  [1]{#1}%
\providecommand \bibfnamefont [1]{#1}%
\providecommand \citenamefont [1]{#1}%
\providecommand \href@noop [0]{\@secondoftwo}%
\providecommand \href [0]{\begingroup \@sanitize@url \@href}%
\providecommand \@href[1]{\@@startlink{#1}\@@href}%
\providecommand \@@href[1]{\endgroup#1\@@endlink}%
\providecommand \@sanitize@url [0]{\catcode `\\12\catcode `\$12\catcode `\&12\catcode `\#12\catcode `\^12\catcode `\_12\catcode `\%12\relax}%
\providecommand \@@startlink[1]{}%
\providecommand \@@endlink[0]{}%
\providecommand \url  [0]{\begingroup\@sanitize@url \@url }%
\providecommand \@url [1]{\endgroup\@href {#1}{\urlprefix }}%
\providecommand \urlprefix  [0]{URL }%
\providecommand \Eprint [0]{\href }%
\providecommand \doibase [0]{http://dx.doi.org/}%
\providecommand \selectlanguage [0]{\@gobble}%
\providecommand \bibinfo  [0]{\@secondoftwo}%
\providecommand \bibfield  [0]{\@secondoftwo}%
\providecommand \translation [1]{[#1]}%
\providecommand \BibitemOpen [0]{}%
\providecommand \bibitemStop [0]{}%
\providecommand \bibitemNoStop [0]{.\EOS\space}%
\providecommand \EOS [0]{\spacefactor3000\relax}%
\providecommand \BibitemShut  [1]{\csname bibitem#1\endcsname}%
\let\auto@bib@innerbib\@empty
\bibitem [{\citenamefont {Lyons}\ \emph {et~al.}(2019)\citenamefont {Lyons} \emph {et~al.}}]{Lyons2019}%
  \BibitemOpen
  \bibfield  {author} {\bibinfo {author} {\bibfnamefont {S.}~\bibnamefont {Lyons}} \emph {et~al.},\ }\bibfield  {title} {\enquote {\bibinfo {title} {$^{69,71}\mathrm{Co}~ \ensuremath{\beta}$-decay strength distributions from total absorption spectroscopy},}\ }\href {\doibase 10.1103/PhysRevC.100.025806} {\bibfield  {journal} {\bibinfo  {journal} {Phys. Rev. C}\ }\textbf {\bibinfo {volume} {100}},\ \bibinfo {pages} {025806} (\bibinfo {year} {2019})}\BibitemShut {NoStop}%
\bibitem [{\citenamefont {Dombos}\ \emph {et~al.}(2019)\citenamefont {Dombos} \emph {et~al.}}]{Dom19}%
  \BibitemOpen
  \bibfield  {author} {\bibinfo {author} {\bibfnamefont {A.~C.}\ \bibnamefont {Dombos}} \emph {et~al.},\ }\bibfield  {title} {\enquote {\bibinfo {title} {$\ensuremath{\beta}$-decay half-lives of neutron-rich nuclides in the ${A}=100-110$ mass region},}\ }\href {\doibase 10.1103/PhysRevC.99.015802} {\bibfield  {journal} {\bibinfo  {journal} {Phys. Rev. C}\ }\textbf {\bibinfo {volume} {99}},\ \bibinfo {pages} {015802} (\bibinfo {year} {2019})}\BibitemShut {NoStop}%
\bibitem [{\citenamefont {Morrissey}\ \emph {et~al.}(2003)\citenamefont {Morrissey}, \citenamefont {Sherrill}, \citenamefont {Steiner}, \citenamefont {Stolz},\ and\ \citenamefont {Wiedenhoever}}]{Mor03}%
  \BibitemOpen
  \bibfield  {author} {\bibinfo {author} {\bibfnamefont {D.J.}\ \bibnamefont {Morrissey}}, \bibinfo {author} {\bibfnamefont {B.M.}\ \bibnamefont {Sherrill}}, \bibinfo {author} {\bibfnamefont {M.}~\bibnamefont {Steiner}}, \bibinfo {author} {\bibfnamefont {A.}~\bibnamefont {Stolz}}, \ and\ \bibinfo {author} {\bibfnamefont {I.}~\bibnamefont {Wiedenhoever}},\ }\bibfield  {title} {\enquote {\bibinfo {title} {Commissioning the {A}1900 projectile fragment separator},}\ }\href {\doibase https://doi.org/10.1016/S0168-583X(02)01895-5} {\bibfield  {journal} {\bibinfo  {journal} {Nucl. Instr. Meth. Phys. Res. B}\ }\textbf {\bibinfo {volume} {204}},\ \bibinfo {pages} {90--96} (\bibinfo {year} {2003})}\BibitemShut {NoStop}%
\bibitem [{\citenamefont {Dombos}\ \emph {et~al.}(2016)\citenamefont {Dombos} \emph {et~al.}}]{Dom16}%
  \BibitemOpen
  \bibfield  {author} {\bibinfo {author} {\bibfnamefont {A.~C.}\ \bibnamefont {Dombos}} \emph {et~al.},\ }\bibfield  {title} {\enquote {\bibinfo {title} {Total absorption spectroscopy of the $\ensuremath{\beta}$ decay of $^{76}\mathrm{Ga}$},}\ }\href {\doibase 10.1103/PhysRevC.93.064317} {\bibfield  {journal} {\bibinfo  {journal} {Phys. Rev. C}\ }\textbf {\bibinfo {volume} {93}},\ \bibinfo {pages} {064317} (\bibinfo {year} {2016})}\BibitemShut {NoStop}%
\bibitem [{\citenamefont {Agostinelli}\ \emph {et~al.}(2003)\citenamefont {Agostinelli} \emph {et~al.}}]{Agostinelli2003}%
  \BibitemOpen
  \bibfield  {author} {\bibinfo {author} {\bibfnamefont {S.}~\bibnamefont {Agostinelli}} \emph {et~al.},\ }\bibfield  {title} {\enquote {\bibinfo {title} {Geant4—a simulation toolkit},}\ }\href {\doibase https://doi.org/10.1016/S0168-9002(03)01368-8} {\bibfield  {journal} {\bibinfo  {journal} {Nucl. Instr. Meth. Phys. Res. A}\ }\textbf {\bibinfo {volume} {506}},\ \bibinfo {pages} {250--303} (\bibinfo {year} {2003})}\BibitemShut {NoStop}%
\bibitem [{\citenamefont {Gürdal}\ and\ \citenamefont {McCutchan}(2016)}]{Gurdal2016}%
  \BibitemOpen
  \bibfield  {author} {\bibinfo {author} {\bibfnamefont {G.}~\bibnamefont {Gürdal}}\ and\ \bibinfo {author} {\bibfnamefont {E.A.}\ \bibnamefont {McCutchan}},\ }\bibfield  {title} {\enquote {\bibinfo {title} {Nuclear data sheets for {A} = 70},}\ }\href {\doibase https://doi.org/10.1016/j.nds.2016.08.001} {\bibfield  {journal} {\bibinfo  {journal} {Nucl. Data Sheets}\ }\textbf {\bibinfo {volume} {136}},\ \bibinfo {pages} {1--162} (\bibinfo {year} {2016})}\BibitemShut {NoStop}%
\bibitem [{\citenamefont {Kirsch}\ and\ \citenamefont {Bernstein}(2018)}]{Kirsch2018}%
  \BibitemOpen
  \bibfield  {author} {\bibinfo {author} {\bibfnamefont {L.E.}\ \bibnamefont {Kirsch}}\ and\ \bibinfo {author} {\bibfnamefont {L.A.}\ \bibnamefont {Bernstein}},\ }\bibfield  {title} {\enquote {\bibinfo {title} {R{AINIER}: A simulation tool for distributions of excited nuclear states and cascade fluctuations},}\ }\href {\doibase https://doi.org/10.1016/j.nima.2018.02.096} {\bibfield  {journal} {\bibinfo  {journal} {Nucl. Instr. Meth. Phys. Res. A}\ }\textbf {\bibinfo {volume} {892}},\ \bibinfo {pages} {30--40} (\bibinfo {year} {2018})}\BibitemShut {NoStop}%
\bibitem [{\citenamefont {von Egidy}\ and\ \citenamefont {Bucurescu}(2009)}]{Egi09}%
  \BibitemOpen
  \bibfield  {author} {\bibinfo {author} {\bibfnamefont {T.}~\bibnamefont {von Egidy}}\ and\ \bibinfo {author} {\bibfnamefont {D.}~\bibnamefont {Bucurescu}},\ }\bibfield  {title} {\enquote {\bibinfo {title} {Experimental energy-dependent nuclear spin distributions},}\ }\href {\doibase 10.1103/PhysRevC.80.054310} {\bibfield  {journal} {\bibinfo  {journal} {Phys. Rev. C}\ }\textbf {\bibinfo {volume} {80}},\ \bibinfo {pages} {054310} (\bibinfo {year} {2009})}\BibitemShut {NoStop}%
\bibitem [{\citenamefont {Liddick}\ \emph {et~al.}(2016)\citenamefont {Liddick} \emph {et~al.}}]{Lid16}%
  \BibitemOpen
  \bibfield  {author} {\bibinfo {author} {\bibfnamefont {S.~N.}\ \bibnamefont {Liddick}} \emph {et~al.},\ }\bibfield  {title} {\enquote {\bibinfo {title} {Experimental neutron capture rate constraint far from stability},}\ }\href {\doibase 10.1103/PhysRevLett.116.242502} {\bibfield  {journal} {\bibinfo  {journal} {Phys. Rev. Lett.}\ }\textbf {\bibinfo {volume} {116}},\ \bibinfo {pages} {242502} (\bibinfo {year} {2016})}\BibitemShut {NoStop}%
\bibitem [{\citenamefont {Larsen}\ \emph {et~al.}(2018)\citenamefont {Larsen} \emph {et~al.}}]{Lar18}%
  \BibitemOpen
  \bibfield  {author} {\bibinfo {author} {\bibfnamefont {A.~C.}\ \bibnamefont {Larsen}} \emph {et~al.},\ }\bibfield  {title} {\enquote {\bibinfo {title} {Enhanced low-energy $\ensuremath{\gamma}\text{-decay}$ strength of $^{70}\mathrm{Ni}$ and its robustness within the shell model},}\ }\href {\doibase 10.1103/PhysRevC.97.054329} {\bibfield  {journal} {\bibinfo  {journal} {Phys. Rev. C}\ }\textbf {\bibinfo {volume} {97}},\ \bibinfo {pages} {054329} (\bibinfo {year} {2018})}\BibitemShut {NoStop}%
\bibitem [{\citenamefont {Hardy}\ and\ \citenamefont {Towner}(2009)}]{Hardy2009}%
  \BibitemOpen
  \bibfield  {author} {\bibinfo {author} {\bibfnamefont {J.~C.}\ \bibnamefont {Hardy}}\ and\ \bibinfo {author} {\bibfnamefont {I.~S.}\ \bibnamefont {Towner}},\ }\bibfield  {title} {\enquote {\bibinfo {title} {Superallowed ${0}^{+}\ensuremath{\rightarrow}{0}^{+}$ nuclear $\ensuremath{\beta}$ decays: A new survey with precision tests of the conserved vector current hypothesis and the standard model},}\ }\href {\doibase 10.1103/PhysRevC.79.055502} {\bibfield  {journal} {\bibinfo  {journal} {Phys. Rev. C}\ }\textbf {\bibinfo {volume} {79}},\ \bibinfo {pages} {055502} (\bibinfo {year} {2009})}\BibitemShut {NoStop}%
\bibitem [{\citenamefont {Yao}(2006)}]{Yao2006}%
  \BibitemOpen
  \bibfield  {author} {\bibinfo {author} {\bibfnamefont {W-M.}\ \bibnamefont {Yao}},\ }\bibfield  {title} {\enquote {\bibinfo {title} {Review of particle physics},}\ }\href {\doibase 10.1088/0954-3899/33/1/001} {\bibfield  {journal} {\bibinfo  {journal} {J. Phys. G}\ }\textbf {\bibinfo {volume} {33}},\ \bibinfo {pages} {1} (\bibinfo {year} {2006})}\BibitemShut {NoStop}%
\bibitem [{\citenamefont {Hjorth-Jensen}\ \emph {et~al.}(1995)\citenamefont {Hjorth-Jensen}, \citenamefont {Kuo},\ and\ \citenamefont {Osnes}}]{jensen}%
  \BibitemOpen
  \bibfield  {author} {\bibinfo {author} {\bibfnamefont {M.}~\bibnamefont {Hjorth-Jensen}}, \bibinfo {author} {\bibfnamefont {T.~T.~S.}\ \bibnamefont {Kuo}}, \ and\ \bibinfo {author} {\bibfnamefont {E.}~\bibnamefont {Osnes}},\ }\bibfield  {title} {\enquote {\bibinfo {title} {Realistic effective interactions for nuclear systems},}\ }\href {\doibase https://doi.org/10.1016/0370-1573(95)00012-6} {\bibfield  {journal} {\bibinfo  {journal} {Phys. Rep.}\ }\textbf {\bibinfo {volume} {261}},\ \bibinfo {pages} {125--270} (\bibinfo {year} {1995})}\BibitemShut {NoStop}%
\bibitem [{\citenamefont {Bogner}\ \emph {et~al.}(2003)\citenamefont {Bogner}, \citenamefont {Kuo},\ and\ \citenamefont {Schwenk}}]{Bogner2003}%
  \BibitemOpen
  \bibfield  {author} {\bibinfo {author} {\bibfnamefont {S.K.}\ \bibnamefont {Bogner}}, \bibinfo {author} {\bibfnamefont {T.T.S.}\ \bibnamefont {Kuo}}, \ and\ \bibinfo {author} {\bibfnamefont {A.}~\bibnamefont {Schwenk}},\ }\bibfield  {title} {\enquote {\bibinfo {title} {Model-independent low momentum nucleon interaction from phase shift equivalence},}\ }\href {\doibase https://doi.org/10.1016/j.physrep.2003.07.001} {\bibfield  {journal} {\bibinfo  {journal} {Phys. Rep.}\ }\textbf {\bibinfo {volume} {386}},\ \bibinfo {pages} {1--27} (\bibinfo {year} {2003})}\BibitemShut {NoStop}%
\bibitem [{\citenamefont {Lisetskiy}\ \emph {et~al.}(2004)\citenamefont {Lisetskiy}, \citenamefont {Brown}, \citenamefont {Horoi},\ and\ \citenamefont {Grawe}}]{z28}%
  \BibitemOpen
  \bibfield  {author} {\bibinfo {author} {\bibfnamefont {A.~F.}\ \bibnamefont {Lisetskiy}}, \bibinfo {author} {\bibfnamefont {B.~A.}\ \bibnamefont {Brown}}, \bibinfo {author} {\bibfnamefont {M.}~\bibnamefont {Horoi}}, \ and\ \bibinfo {author} {\bibfnamefont {H.}~\bibnamefont {Grawe}},\ }\bibfield  {title} {\enquote {\bibinfo {title} {New ${T}=1$ effective interactions for the ${f}_{5∕2}\phantom{\rule{0.3em}{0ex}}{p}_{3∕2}\phantom{\rule{0.3em}{0ex}}{p}_{1∕2}\phantom{\rule{0.3em}{0ex}}{g}_{9∕2}$ model space: Implications for valence-mirror symmetry and seniority isomers},}\ }\href {\doibase 10.1103/PhysRevC.70.044314} {\bibfield  {journal} {\bibinfo  {journal} {Phys. Rev. C}\ }\textbf {\bibinfo {volume} {70}},\ \bibinfo {pages} {044314} (\bibinfo {year} {2004})}\BibitemShut {NoStop}%
\bibitem [{\citenamefont {Mukhopadhyay}\ \emph {et~al.}(2017)\citenamefont {Mukhopadhyay} \emph {et~al.}}]{jj44a}%
  \BibitemOpen
  \bibfield  {author} {\bibinfo {author} {\bibfnamefont {S.}~\bibnamefont {Mukhopadhyay}} \emph {et~al.},\ }\bibfield  {title} {\enquote {\bibinfo {title} {Nuclear structure of $^{76}\mathrm{Ge}$ from inelastic neutron scattering measurements and shell model calculations},}\ }\href {\doibase 10.1103/PhysRevC.95.014327} {\bibfield  {journal} {\bibinfo  {journal} {Phys. Rev. C}\ }\textbf {\bibinfo {volume} {95}},\ \bibinfo {pages} {014327} (\bibinfo {year} {2017})}\BibitemShut {NoStop}%
\bibitem [{\citenamefont {Brown}\ and\ \citenamefont {Wildenthal}(1988)}]{rds}%
  \BibitemOpen
  \bibfield  {author} {\bibinfo {author} {\bibfnamefont {B~A}\ \bibnamefont {Brown}}\ and\ \bibinfo {author} {\bibfnamefont {B~H}\ \bibnamefont {Wildenthal}},\ }\bibfield  {title} {\enquote {\bibinfo {title} {Status of the nuclear shell model},}\ }\href {\doibase 10.1146/annurev.ns.38.120188.000333} {\bibfield  {journal} {\bibinfo  {journal} {Annu. Rev. Nucl. Part. Sci.}\ }\textbf {\bibinfo {volume} {38}},\ \bibinfo {pages} {29--66} (\bibinfo {year} {1988})}\BibitemShut {NoStop}%
\bibitem [{\citenamefont {Caurier}\ \emph {et~al.}(1994)\citenamefont {Caurier}, \citenamefont {Zuker}, \citenamefont {Poves},\ and\ \citenamefont {Mart\'{\i}nez-Pinedo}}]{rfp}%
  \BibitemOpen
  \bibfield  {author} {\bibinfo {author} {\bibfnamefont {E.}~\bibnamefont {Caurier}}, \bibinfo {author} {\bibfnamefont {A.~P.}\ \bibnamefont {Zuker}}, \bibinfo {author} {\bibfnamefont {A.}~\bibnamefont {Poves}}, \ and\ \bibinfo {author} {\bibfnamefont {G.}~\bibnamefont {Mart\'{\i}nez-Pinedo}},\ }\bibfield  {title} {\enquote {\bibinfo {title} {Full pf shell model study of {A}=48 nuclei},}\ }\href {\doibase 10.1103/PhysRevC.50.225} {\bibfield  {journal} {\bibinfo  {journal} {Phys. Rev. C}\ }\textbf {\bibinfo {volume} {50}},\ \bibinfo {pages} {225--236} (\bibinfo {year} {1994})}\BibitemShut {NoStop}%
\bibitem [{\citenamefont {Goodman}\ \emph {et~al.}(1990)\citenamefont {Goodman}, \citenamefont {Rapaport},\ and\ \citenamefont {Bloom}}]{pn}%
  \BibitemOpen
  \bibfield  {author} {\bibinfo {author} {\bibfnamefont {C.~D.}\ \bibnamefont {Goodman}}, \bibinfo {author} {\bibfnamefont {J.}~\bibnamefont {Rapaport}}, \ and\ \bibinfo {author} {\bibfnamefont {S.~D.}\ \bibnamefont {Bloom}},\ }\bibfield  {title} {\enquote {\bibinfo {title} {Comment on ``testing'' the gamow-teller sum rule},}\ }\href {\doibase 10.1103/PhysRevC.42.1150} {\bibfield  {journal} {\bibinfo  {journal} {Phys. Rev. C}\ }\textbf {\bibinfo {volume} {42}},\ \bibinfo {pages} {1150--1152} (\bibinfo {year} {1990})}\BibitemShut {NoStop}%
\bibitem [{\citenamefont {Nilsson}(1955)}]{Nilsson1955}%
  \BibitemOpen
  \bibfield  {author} {\bibinfo {author} {\bibfnamefont {S~G}\ \bibnamefont {Nilsson}},\ }\bibfield  {title} {\enquote {\bibinfo {title} {Binding states of individual nucleons in strongly deformed nuclei},}\ }\href {https://www.osti.gov/biblio/4376487} {\bibfield  {journal} {\bibinfo  {journal} {Kgl. Danske Videnskab. Selskab., Mat.-fys Medd.}\ }\textbf {\bibinfo {volume} {Vol: 29, No. 16}} (\bibinfo {year} {1955})}\BibitemShut {NoStop}%
\bibitem [{\citenamefont {Mueller}\ \emph {et~al.}(1999)\citenamefont {Mueller} \emph {et~al.}}]{Mue99}%
  \BibitemOpen
  \bibfield  {author} {\bibinfo {author} {\bibfnamefont {W.~F.}\ \bibnamefont {Mueller}} \emph {et~al.},\ }\bibfield  {title} {\enquote {\bibinfo {title} {Magicity of the $^{68}${N}i semidouble-closed-shell nucleus probed by gamow-teller decay of the odd-$\mathit{A}$ neighbors},}\ }\href {\doibase 10.1103/PhysRevLett.83.3613} {\bibfield  {journal} {\bibinfo  {journal} {Phys. Rev. Lett.}\ }\textbf {\bibinfo {volume} {83}},\ \bibinfo {pages} {3613--3616} (\bibinfo {year} {1999})}\BibitemShut {NoStop}%
\bibitem [{\citenamefont {Liddick}\ \emph {et~al.}(2015)\citenamefont {Liddick} \emph {et~al.}}]{lid15}%
  \BibitemOpen
  \bibfield  {author} {\bibinfo {author} {\bibfnamefont {S.~N.}\ \bibnamefont {Liddick}} \emph {et~al.},\ }\bibfield  {title} {\enquote {\bibinfo {title} {Analogous intruder behavior near {N}i, {S}n, and {P}b isotopes},}\ }\href {\doibase 10.1103/PhysRevC.92.024319} {\bibfield  {journal} {\bibinfo  {journal} {Phys. Rev. C}\ }\textbf {\bibinfo {volume} {92}},\ \bibinfo {pages} {024319} (\bibinfo {year} {2015})}\BibitemShut {NoStop}%
\bibitem [{\citenamefont {Lokotko}\ \emph {et~al.}(2020)\citenamefont {Lokotko} \emph {et~al.}}]{Lokotko2020}%
  \BibitemOpen
  \bibfield  {author} {\bibinfo {author} {\bibfnamefont {T.}~\bibnamefont {Lokotko}} \emph {et~al.},\ }\bibfield  {title} {\enquote {\bibinfo {title} {Shell structure of the neutron-rich isotopes $^{69,71,73}\mathrm{Co}$},}\ }\href {\doibase 10.1103/PhysRevC.101.034314} {\bibfield  {journal} {\bibinfo  {journal} {Phys. Rev. C}\ }\textbf {\bibinfo {volume} {101}},\ \bibinfo {pages} {034314} (\bibinfo {year} {2020})}\BibitemShut {NoStop}%
\bibitem [{\citenamefont {Longfellow}\ \emph {et~al.}(2020)\citenamefont {Longfellow} \emph {et~al.}}]{Longfellow2020}%
  \BibitemOpen
  \bibfield  {author} {\bibinfo {author} {\bibfnamefont {B.}~\bibnamefont {Longfellow}} \emph {et~al.},\ }\bibfield  {title} {\enquote {\bibinfo {title} {Shape changes in the ${N}=28$ island of inversion: Collective structures built on configuration-coexisting states in $^{43}\mathrm{S}$},}\ }\href {\doibase 10.1103/PhysRevLett.125.232501} {\bibfield  {journal} {\bibinfo  {journal} {Phys. Rev. Lett.}\ }\textbf {\bibinfo {volume} {125}},\ \bibinfo {pages} {232501} (\bibinfo {year} {2020})}\BibitemShut {NoStop}%
\bibitem [{\citenamefont {Cwiok}\ \emph {et~al.}(1987)\citenamefont {Cwiok}, \citenamefont {Dudek}, \citenamefont {Nazarewicz}, \citenamefont {Skalski},\ and\ \citenamefont {Werner}}]{dws}%
  \BibitemOpen
  \bibfield  {author} {\bibinfo {author} {\bibfnamefont {S.}~\bibnamefont {Cwiok}}, \bibinfo {author} {\bibfnamefont {J.}~\bibnamefont {Dudek}}, \bibinfo {author} {\bibfnamefont {W.}~\bibnamefont {Nazarewicz}}, \bibinfo {author} {\bibfnamefont {J.}~\bibnamefont {Skalski}}, \ and\ \bibinfo {author} {\bibfnamefont {T.}~\bibnamefont {Werner}},\ }\bibfield  {title} {\enquote {\bibinfo {title} {Single-particle energies, wave functions, quadrupole moments and g-factors in an axially deformed woods-saxon potential with applications to the two-centre-type nuclear problems},}\ }\href {\doibase https://doi.org/10.1016/0010-4655(87)90093-2} {\bibfield  {journal} {\bibinfo  {journal} {Comput. Phys. Commun.}\ }\textbf {\bibinfo {volume} {46}},\ \bibinfo {pages} {379--399} (\bibinfo {year} {1987})}\BibitemShut {NoStop}%
\bibitem [{\citenamefont {Chakrabarti}\ \emph {et~al.}(2024)\citenamefont {Chakrabarti} \emph {et~al.}}]{cr51}%
  \BibitemOpen
  \bibfield  {author} {\bibinfo {author} {\bibfnamefont {R.}~\bibnamefont {Chakrabarti}} \emph {et~al.},\ }\bibfield  {title} {\enquote {\bibinfo {title} {Extended level structure of $^{51}\mathrm{Cr}$ with measured mean lifetimes of yrast states in agreement with shell-model calculations},}\ }\href {\doibase 10.1103/PhysRevC.109.064311} {\bibfield  {journal} {\bibinfo  {journal} {Phys. Rev. C}\ }\textbf {\bibinfo {volume} {109}},\ \bibinfo {pages} {064311} (\bibinfo {year} {2024})}\BibitemShut {NoStop}%
\bibitem [{\citenamefont {Heyde}\ \emph {et~al.}(1983)\citenamefont {Heyde}, \citenamefont {{Van Isacker}}, \citenamefont {Waroquier}, \citenamefont {Wood},\ and\ \citenamefont {Meyer}}]{Heyde1983}%
  \BibitemOpen
  \bibfield  {author} {\bibinfo {author} {\bibfnamefont {K.}~\bibnamefont {Heyde}}, \bibinfo {author} {\bibfnamefont {P.}~\bibnamefont {{Van Isacker}}}, \bibinfo {author} {\bibfnamefont {M.}~\bibnamefont {Waroquier}}, \bibinfo {author} {\bibfnamefont {J.L.}\ \bibnamefont {Wood}}, \ and\ \bibinfo {author} {\bibfnamefont {R.A.}\ \bibnamefont {Meyer}},\ }\bibfield  {title} {\enquote {\bibinfo {title} {Coexistence in odd-mass nuclei},}\ }\href {\doibase https://doi.org/10.1016/0370-1573(83)90085-6} {\bibfield  {journal} {\bibinfo  {journal} {Phys. Rep.}\ }\textbf {\bibinfo {volume} {102}},\ \bibinfo {pages} {291--393} (\bibinfo {year} {1983})}\BibitemShut {NoStop}%
\bibitem [{\citenamefont {Honma}\ \emph {et~al.}(2004)\citenamefont {Honma}, \citenamefont {Otsuka}, \citenamefont {Brown},\ and\ \citenamefont {Mizusaki}}]{gx1a}%
  \BibitemOpen
  \bibfield  {author} {\bibinfo {author} {\bibfnamefont {M.}~\bibnamefont {Honma}}, \bibinfo {author} {\bibfnamefont {T.}~\bibnamefont {Otsuka}}, \bibinfo {author} {\bibfnamefont {B.~A.}\ \bibnamefont {Brown}}, \ and\ \bibinfo {author} {\bibfnamefont {T.}~\bibnamefont {Mizusaki}},\ }\bibfield  {title} {\enquote {\bibinfo {title} {New effective interaction for $pf$-shell nuclei and its implications for the stability of the ${N}={Z}=28$ closed core},}\ }\href {\doibase 10.1103/PhysRevC.69.034335} {\bibfield  {journal} {\bibinfo  {journal} {Phys. Rev. C}\ }\textbf {\bibinfo {volume} {69}},\ \bibinfo {pages} {034335} (\bibinfo {year} {2004})}\BibitemShut {NoStop}%
\end{thebibliography}%


\begin{thebibliography}{31}%
\makeatletter
\providecommand \@ifxundefined [1]{%
 \@ifx{#1\undefined}
}%
\providecommand \@ifnum [1]{%
 \ifnum #1\expandafter \@firstoftwo
 \else \expandafter \@secondoftwo
 \fi
}%
\providecommand \@ifx [1]{%
 \ifx #1\expandafter \@firstoftwo
 \else \expandafter \@secondoftwo
 \fi
}%
\providecommand \natexlab [1]{#1}%
\providecommand \enquote  [1]{``#1''}%
\providecommand \bibnamefont  [1]{#1}%
\providecommand \bibfnamefont [1]{#1}%
\providecommand \citenamefont [1]{#1}%
\providecommand \href@noop [0]{\@secondoftwo}%
\providecommand \href [0]{\begingroup \@sanitize@url \@href}%
\providecommand \@href[1]{\@@startlink{#1}\@@href}%
\providecommand \@@href[1]{\endgroup#1\@@endlink}%
\providecommand \@sanitize@url [0]{\catcode `\\12\catcode `\$12\catcode `\&12\catcode `\#12\catcode `\^12\catcode `\_12\catcode `\%12\relax}%
\providecommand \@@startlink[1]{}%
\providecommand \@@endlink[0]{}%
\providecommand \url  [0]{\begingroup\@sanitize@url \@url }%
\providecommand \@url [1]{\endgroup\@href {#1}{\urlprefix }}%
\providecommand \urlprefix  [0]{URL }%
\providecommand \Eprint [0]{\href }%
\providecommand \doibase [0]{http://dx.doi.org/}%
\providecommand \selectlanguage [0]{\@gobble}%
\providecommand \bibinfo  [0]{\@secondoftwo}%
\providecommand \bibfield  [0]{\@secondoftwo}%
\providecommand \translation [1]{[#1]}%
\providecommand \BibitemOpen [0]{}%
\providecommand \bibitemStop [0]{}%
\providecommand \bibitemNoStop [0]{.\EOS\space}%
\providecommand \EOS [0]{\spacefactor3000\relax}%
\providecommand \BibitemShut  [1]{\csname bibitem#1\endcsname}%
\let\auto@bib@innerbib\@empty
\bibitem [{\citenamefont {Mayer}(1949)}]{may45}%
  \BibitemOpen
  \bibfield  {author} {\bibinfo {author} {\bibfnamefont {M.~G.}\ \bibnamefont {Mayer}},\ }\bibfield  {title} {\enquote {\bibinfo {title} {On closed shells in nuclei. {II}},}\ }\href {\doibase 10.1103/PhysRev.75.1969} {\bibfield  {journal} {\bibinfo  {journal} {Phys. Rev.}\ }\textbf {\bibinfo {volume} {75}},\ \bibinfo {pages} {1969--1970} (\bibinfo {year} {1949})}\BibitemShut {NoStop}%
\bibitem [{\citenamefont {Haxel}\ \emph {et~al.}(1949)\citenamefont {Haxel}, \citenamefont {Jensen},\ and\ \citenamefont {Suess}}]{hax49}%
  \BibitemOpen
  \bibfield  {author} {\bibinfo {author} {\bibfnamefont {O.}~\bibnamefont {Haxel}}, \bibinfo {author} {\bibfnamefont {J.~H.~D.}\ \bibnamefont {Jensen}}, \ and\ \bibinfo {author} {\bibfnamefont {H.~E.}\ \bibnamefont {Suess}},\ }\bibfield  {title} {\enquote {\bibinfo {title} {On the "magic numbers" in nuclear structure},}\ }\href {\doibase 10.1103/PhysRev.75.1766.2} {\bibfield  {journal} {\bibinfo  {journal} {Phys. Rev.}\ }\textbf {\bibinfo {volume} {75}},\ \bibinfo {pages} {1766--1766} (\bibinfo {year} {1949})}\BibitemShut {NoStop}%
\bibitem [{\citenamefont {Bohr}(1913)}]{boh13}%
  \BibitemOpen
  \bibfield  {author} {\bibinfo {author} {\bibfnamefont {N.}~\bibnamefont {Bohr}},\ }\bibfield  {title} {\enquote {\bibinfo {title} {I. on the constitution of atoms and molecules},}\ }\href {\doibase 10.1080/14786441308634955} {\bibfield  {journal} {\bibinfo  {journal} {Lond. Edinb. Dubl. Phil. Mag.}\ }\textbf {\bibinfo {volume} {26}},\ \bibinfo {pages} {1--25} (\bibinfo {year} {1913})}\BibitemShut {NoStop}%
\bibitem [{\citenamefont {Tarucha}\ \emph {et~al.}(1996)\citenamefont {Tarucha}, \citenamefont {Austing}, \citenamefont {Honda}, \citenamefont {van~der Hage},\ and\ \citenamefont {Kouwenhoven}}]{tar96}%
  \BibitemOpen
  \bibfield  {author} {\bibinfo {author} {\bibfnamefont {S.}~\bibnamefont {Tarucha}}, \bibinfo {author} {\bibfnamefont {D.~G.}\ \bibnamefont {Austing}}, \bibinfo {author} {\bibfnamefont {T.}~\bibnamefont {Honda}}, \bibinfo {author} {\bibfnamefont {R.~J.}\ \bibnamefont {van~der Hage}}, \ and\ \bibinfo {author} {\bibfnamefont {L.~P.}\ \bibnamefont {Kouwenhoven}},\ }\bibfield  {title} {\enquote {\bibinfo {title} {Shell filling and spin effects in a few electron quantum dot},}\ }\href {\doibase 10.1103/PhysRevLett.77.3613} {\bibfield  {journal} {\bibinfo  {journal} {Phys. Rev. Lett.}\ }\textbf {\bibinfo {volume} {77}},\ \bibinfo {pages} {3613--3616} (\bibinfo {year} {1996})}\BibitemShut {NoStop}%
\bibitem [{\citenamefont {Knight}\ \emph {et~al.}(1984)\citenamefont {Knight} \emph {et~al.}}]{kni84}%
  \BibitemOpen
  \bibfield  {author} {\bibinfo {author} {\bibfnamefont {W.~D.}\ \bibnamefont {Knight}} \emph {et~al.},\ }\bibfield  {title} {\enquote {\bibinfo {title} {Electronic shell structure and abundances of sodium clusters},}\ }\href {\doibase 10.1103/PhysRevLett.52.2141} {\bibfield  {journal} {\bibinfo  {journal} {Phys. Rev. Lett.}\ }\textbf {\bibinfo {volume} {52}},\ \bibinfo {pages} {2141--2143} (\bibinfo {year} {1984})}\BibitemShut {NoStop}%
\bibitem [{\citenamefont {Morinaga}(1956)}]{mor56}%
  \BibitemOpen
  \bibfield  {author} {\bibinfo {author} {\bibfnamefont {H.}~\bibnamefont {Morinaga}},\ }\bibfield  {title} {\enquote {\bibinfo {title} {Interpretation of some of the excited states of $4n$ self-conjugate nuclei},}\ }\href {\doibase 10.1103/PhysRev.101.254} {\bibfield  {journal} {\bibinfo  {journal} {Phys. Rev.}\ }\textbf {\bibinfo {volume} {101}},\ \bibinfo {pages} {254--258} (\bibinfo {year} {1956})}\BibitemShut {NoStop}%
\bibitem [{\citenamefont {Garrett}\ \emph {et~al.}(2022)\citenamefont {Garrett}, \citenamefont {Zielińska},\ and\ \citenamefont {Clément}}]{gar22}%
  \BibitemOpen
  \bibfield  {author} {\bibinfo {author} {\bibfnamefont {P.~E.}\ \bibnamefont {Garrett}}, \bibinfo {author} {\bibfnamefont {M.}~\bibnamefont {Zielińska}}, \ and\ \bibinfo {author} {\bibfnamefont {E.}~\bibnamefont {Clément}},\ }\bibfield  {title} {\enquote {\bibinfo {title} {An experimental view on shape coexistence in nuclei},}\ }\href {\doibase https://doi.org/10.1016/j.ppnp.2021.103931} {\bibfield  {journal} {\bibinfo  {journal} {Prog. Part. Nucl. Phys.}\ }\textbf {\bibinfo {volume} {124}},\ \bibinfo {pages} {103931} (\bibinfo {year} {2022})}\BibitemShut {NoStop}%
\bibitem [{\citenamefont {Gaffney}\ \emph {et~al.}(2013)\citenamefont {Gaffney} \emph {et~al.}}]{gaf13}%
  \BibitemOpen
  \bibfield  {author} {\bibinfo {author} {\bibfnamefont {L.~P.}\ \bibnamefont {Gaffney}} \emph {et~al.},\ }\bibfield  {title} {\enquote {\bibinfo {title} {Studies of pear-shaped nuclei using accelerated radioactive beams},}\ }\href {\doibase https://doi.org/10.1038/nature12073} {\bibfield  {journal} {\bibinfo  {journal} {Nature}\ }\textbf {\bibinfo {volume} {497}},\ \bibinfo {pages} {199--204} (\bibinfo {year} {2013})}\BibitemShut {NoStop}%
\bibitem [{\citenamefont {Taniuchi}\ \emph {et~al.}(2019)\citenamefont {Taniuchi} \emph {et~al.}}]{tan19}%
  \BibitemOpen
  \bibfield  {author} {\bibinfo {author} {\bibfnamefont {R.}~\bibnamefont {Taniuchi}} \emph {et~al.},\ }\bibfield  {title} {\enquote {\bibinfo {title} {$^{78}${N}i revealed as a doubly magic stronghold against nuclear deformation},}\ }\href {\doibase https://doi.org/10.1038/s41586-019-1155-x} {\bibfield  {journal} {\bibinfo  {journal} {Nature}\ }\textbf {\bibinfo {volume} {569}},\ \bibinfo {pages} {53--58} (\bibinfo {year} {2019})}\BibitemShut {NoStop}%
\bibitem [{\citenamefont {Ichikawa}\ \emph {et~al.}(2019)\citenamefont {Ichikawa} \emph {et~al.}}]{ich19}%
  \BibitemOpen
  \bibfield  {author} {\bibinfo {author} {\bibfnamefont {Y.}~\bibnamefont {Ichikawa}} \emph {et~al.},\ }\bibfield  {title} {\enquote {\bibinfo {title} {Interplay between nuclear shell evolution and shape deformation revealed by the magnetic moment of $^{75}${C}u},}\ }\href {\doibase https://doi.org/10.1038/s41567-018-0410-7} {\bibfield  {journal} {\bibinfo  {journal} {Nat. Phys.}\ }\textbf {\bibinfo {volume} {15}},\ \bibinfo {pages} {321--325} (\bibinfo {year} {2019})}\BibitemShut {NoStop}%
\bibitem [{\citenamefont {Reponen}\ \emph {et~al.}(2021)\citenamefont {Reponen} \emph {et~al.}}]{rep21}%
  \BibitemOpen
  \bibfield  {author} {\bibinfo {author} {\bibfnamefont {M.}~\bibnamefont {Reponen}} \emph {et~al.},\ }\bibfield  {title} {\enquote {\bibinfo {title} {Evidence of a sudden increase in the nuclear size of proton-rich silver-96},}\ }\href {\doibase https://doi.org/10.1038/s41467-021-24888-x} {\bibfield  {journal} {\bibinfo  {journal} {Nat. Commun.}\ }\textbf {\bibinfo {volume} {12}},\ \bibinfo {pages} {4596} (\bibinfo {year} {2021})}\BibitemShut {NoStop}%
\bibitem [{\citenamefont {Tsunoda}\ \emph {et~al.}(2020)\citenamefont {Tsunoda} \emph {et~al.}}]{tsu20}%
  \BibitemOpen
  \bibfield  {author} {\bibinfo {author} {\bibfnamefont {N.}~\bibnamefont {Tsunoda}} \emph {et~al.},\ }\bibfield  {title} {\enquote {\bibinfo {title} {The impact of nuclear shape on the emergence of the neutron dripline},}\ }\href {\doibase https://doi.org/10.1038/s41586-020-2848-x} {\bibfield  {journal} {\bibinfo  {journal} {Nature}\ }\textbf {\bibinfo {volume} {587}},\ \bibinfo {pages} {66--71} (\bibinfo {year} {2020})}\BibitemShut {NoStop}%
\bibitem [{\citenamefont {Smits}\ \emph {et~al.}(2024)\citenamefont {Smits}, \citenamefont {Düllmann}, \citenamefont {Indelicato}, \citenamefont {Nazarewicz},\ and\ \citenamefont {Schwerdtfeger}}]{smi23}%
  \BibitemOpen
  \bibfield  {author} {\bibinfo {author} {\bibfnamefont {O.R.}\ \bibnamefont {Smits}}, \bibinfo {author} {\bibfnamefont {C.~E.}\ \bibnamefont {Düllmann}}, \bibinfo {author} {\bibfnamefont {P.}~\bibnamefont {Indelicato}}, \bibinfo {author} {\bibfnamefont {W.}~\bibnamefont {Nazarewicz}}, \ and\ \bibinfo {author} {\bibfnamefont {P.}~\bibnamefont {Schwerdtfeger}},\ }\bibfield  {title} {\enquote {\bibinfo {title} {The quest for superheavy elements and the limit of the periodic table},}\ }\href {\doibase https://doi.org/10.1038/s42254-023-00668-y} {\bibfield  {journal} {\bibinfo  {journal} {Nat. Rev. Phys.}\ }\textbf {\bibinfo {volume} {6}},\ \bibinfo {pages} {86--98} (\bibinfo {year} {2024})}\BibitemShut {NoStop}%
\bibitem [{\citenamefont {Crider}\ \emph {et~al.}(2016)\citenamefont {Crider} \emph {et~al.}}]{Crider2018}%
  \BibitemOpen
  \bibfield  {author} {\bibinfo {author} {\bibfnamefont {B.P.}\ \bibnamefont {Crider}} \emph {et~al.},\ }\bibfield  {title} {\enquote {\bibinfo {title} {Shape coexistence from lifetime and branching-ratio measurements in $^{68,70}${N}i},}\ }\href {\doibase https://doi.org/10.1016/j.physletb.2016.10.020} {\bibfield  {journal} {\bibinfo  {journal} {Phys. Lett. B}\ }\textbf {\bibinfo {volume} {763}},\ \bibinfo {pages} {108--113} (\bibinfo {year} {2016})}\BibitemShut {NoStop}%
\bibitem [{\citenamefont {Prokop}\ \emph {et~al.}(2015)\citenamefont {Prokop} \emph {et~al.}}]{Prokop2015}%
  \BibitemOpen
  \bibfield  {author} {\bibinfo {author} {\bibfnamefont {C.~J.}\ \bibnamefont {Prokop}} \emph {et~al.},\ }\bibfield  {title} {\enquote {\bibinfo {title} {New low-energy ${0}^{+}$ state and shape coexistence in $^{70}\mathrm{Ni}$},}\ }\href {\doibase 10.1103/PhysRevC.92.061302} {\bibfield  {journal} {\bibinfo  {journal} {Phys. Rev. C}\ }\textbf {\bibinfo {volume} {92}},\ \bibinfo {pages} {061302} (\bibinfo {year} {2015})}\BibitemShut {NoStop}%
\bibitem [{\citenamefont {Otsuka}\ and\ \citenamefont {Tsunoda}(2016)}]{Otsuka2016}%
  \BibitemOpen
  \bibfield  {author} {\bibinfo {author} {\bibfnamefont {T.}~\bibnamefont {Otsuka}}\ and\ \bibinfo {author} {\bibfnamefont {Y.}~\bibnamefont {Tsunoda}},\ }\bibfield  {title} {\enquote {\bibinfo {title} {The role of shell evolution in shape coexistence},}\ }\href {\doibase 10.1088/0954-3899/43/2/024009} {\bibfield  {journal} {\bibinfo  {journal} {J. Phys. G}\ }\textbf {\bibinfo {volume} {43}},\ \bibinfo {pages} {024009} (\bibinfo {year} {2016})}\BibitemShut {NoStop}%
\bibitem [{\citenamefont {Morales}\ \emph {et~al.}(2017)\citenamefont {Morales} \emph {et~al.}}]{Mor17}%
  \BibitemOpen
  \bibfield  {author} {\bibinfo {author} {\bibfnamefont {A.I.}\ \bibnamefont {Morales}} \emph {et~al.},\ }\bibfield  {title} {\enquote {\bibinfo {title} {Type {II} shell evolution in {A}=70 isobars from the {N}$\geq$40 island of inversion},}\ }\href {\doibase https://doi.org/10.1016/j.physletb.2016.12.025} {\bibfield  {journal} {\bibinfo  {journal} {Phys. Lett. B}\ }\textbf {\bibinfo {volume} {765}},\ \bibinfo {pages} {328--333} (\bibinfo {year} {2017})}\BibitemShut {NoStop}%
\bibitem [{\citenamefont {Urban}\ \emph {et~al.}(2017)\citenamefont {Urban} \emph {et~al.}}]{Urban2017}%
  \BibitemOpen
  \bibfield  {author} {\bibinfo {author} {\bibfnamefont {W.}~\bibnamefont {Urban}} \emph {et~al.},\ }\bibfield  {title} {\enquote {\bibinfo {title} {Shape coexistence in the odd-odd nucleus $^{98}\mathrm{Y}$: The role of the ${g}_{9/2}$ neutron extruder},}\ }\href {\doibase 10.1103/PhysRevC.96.044333} {\bibfield  {journal} {\bibinfo  {journal} {Phys. Rev. C}\ }\textbf {\bibinfo {volume} {96}},\ \bibinfo {pages} {044333} (\bibinfo {year} {2017})}\BibitemShut {NoStop}%
\bibitem [{\citenamefont {Nowacki}\ \emph {et~al.}(2016)\citenamefont {Nowacki}, \citenamefont {Poves}, \citenamefont {Caurier},\ and\ \citenamefont {Bounthong}}]{Nowacki2016}%
  \BibitemOpen
  \bibfield  {author} {\bibinfo {author} {\bibfnamefont {F.}~\bibnamefont {Nowacki}}, \bibinfo {author} {\bibfnamefont {A.}~\bibnamefont {Poves}}, \bibinfo {author} {\bibfnamefont {E.}~\bibnamefont {Caurier}}, \ and\ \bibinfo {author} {\bibfnamefont {B.}~\bibnamefont {Bounthong}},\ }\bibfield  {title} {\enquote {\bibinfo {title} {Shape coexistence in $^{78}\mathrm{Ni}$ as the portal to the fifth island of inversion},}\ }\href {\doibase 10.1103/PhysRevLett.117.272501} {\bibfield  {journal} {\bibinfo  {journal} {Phys. Rev. Lett.}\ }\textbf {\bibinfo {volume} {117}},\ \bibinfo {pages} {272501} (\bibinfo {year} {2016})}\BibitemShut {NoStop}%
\bibitem [{\citenamefont {Spyrou}\ \emph {et~al.}(2016)\citenamefont {Spyrou} \emph {et~al.}}]{Spy16}%
  \BibitemOpen
  \bibfield  {author} {\bibinfo {author} {\bibfnamefont {A.}~\bibnamefont {Spyrou}} \emph {et~al.},\ }\bibfield  {title} {\enquote {\bibinfo {title} {Strong neutron-$\ensuremath{\gamma}$ competition above the neutron threshold in the decay of $^{70}\mathrm{Co}$},}\ }\href {\doibase 10.1103/PhysRevLett.117.142701} {\bibfield  {journal} {\bibinfo  {journal} {Phys. Rev. Lett.}\ }\textbf {\bibinfo {volume} {117}},\ \bibinfo {pages} {142701} (\bibinfo {year} {2016})}\BibitemShut {NoStop}%
\bibitem [{\citenamefont {Gürdal}\ and\ \citenamefont {McCutchan}(2016)}]{Gurdal2016}%
  \BibitemOpen
  \bibfield  {author} {\bibinfo {author} {\bibfnamefont {G.}~\bibnamefont {Gürdal}}\ and\ \bibinfo {author} {\bibfnamefont {E.A.}\ \bibnamefont {McCutchan}},\ }\bibfield  {title} {\enquote {\bibinfo {title} {Nuclear data sheets for {A} = 70},}\ }\href {\doibase https://doi.org/10.1016/j.nds.2016.08.001} {\bibfield  {journal} {\bibinfo  {journal} {Nucl. Data Sheets}\ }\textbf {\bibinfo {volume} {136}},\ \bibinfo {pages} {1--162} (\bibinfo {year} {2016})}\BibitemShut {NoStop}%
\bibitem [{\citenamefont {Hardy}\ \emph {et~al.}(1977)\citenamefont {Hardy}, \citenamefont {Carraz}, \citenamefont {Jonson},\ and\ \citenamefont {Hansen}}]{har77}%
  \BibitemOpen
  \bibfield  {author} {\bibinfo {author} {\bibfnamefont {J.C.}\ \bibnamefont {Hardy}}, \bibinfo {author} {\bibfnamefont {L.C.}\ \bibnamefont {Carraz}}, \bibinfo {author} {\bibfnamefont {B.}~\bibnamefont {Jonson}}, \ and\ \bibinfo {author} {\bibfnamefont {P.G.}\ \bibnamefont {Hansen}},\ }\bibfield  {title} {\enquote {\bibinfo {title} {The essential decay of pandemonium: A demonstration of errors in complex beta-decay schemes},}\ }\href {\doibase https://doi.org/10.1016/0370-2693(77)90223-4} {\bibfield  {journal} {\bibinfo  {journal} {Phys. Lett. B}\ }\textbf {\bibinfo {volume} {71}},\ \bibinfo {pages} {307--310} (\bibinfo {year} {1977})}\BibitemShut {NoStop}%
\bibitem [{\citenamefont {Sarriguren}\ \emph {et~al.}(2001)\citenamefont {Sarriguren}, \citenamefont {{Moya de Guerra}},\ and\ \citenamefont {Escuderos}}]{Sarriguren2001}%
  \BibitemOpen
  \bibfield  {author} {\bibinfo {author} {\bibfnamefont {P.}~\bibnamefont {Sarriguren}}, \bibinfo {author} {\bibfnamefont {E.}~\bibnamefont {{Moya de Guerra}}}, \ and\ \bibinfo {author} {\bibfnamefont {A.}~\bibnamefont {Escuderos}},\ }\bibfield  {title} {\enquote {\bibinfo {title} {Spin–isospin excitations and $\beta^+$/{EC} half-lives of medium-mass deformed nuclei},}\ }\href {\doibase https://doi.org/10.1016/S0375-9474(01)00565-6} {\bibfield  {journal} {\bibinfo  {journal} {Nucl. Phys. A}\ }\textbf {\bibinfo {volume} {691}},\ \bibinfo {pages} {631--648} (\bibinfo {year} {2001})}\BibitemShut {NoStop}%
\bibitem [{\citenamefont {N\'acher}\ \emph {et~al.}(2004)\citenamefont {N\'acher} \emph {et~al.}}]{Nacher2004}%
  \BibitemOpen
  \bibfield  {author} {\bibinfo {author} {\bibfnamefont {E.}~\bibnamefont {N\'acher}} \emph {et~al.},\ }\bibfield  {title} {\enquote {\bibinfo {title} {Deformation of the ${N}={Z}$ nucleus $^{76}\mathrm{Sr}$ using $\ensuremath{\beta}$-decay studies},}\ }\href {\doibase 10.1103/PhysRevLett.92.232501} {\bibfield  {journal} {\bibinfo  {journal} {Phys. Rev. Lett.}\ }\textbf {\bibinfo {volume} {92}},\ \bibinfo {pages} {232501} (\bibinfo {year} {2004})}\BibitemShut {NoStop}%
\bibitem [{\citenamefont {P\'erez-Cerd\'an}\ \emph {et~al.}(2013)\citenamefont {P\'erez-Cerd\'an} \emph {et~al.}}]{Perez2013}%
  \BibitemOpen
  \bibfield  {author} {\bibinfo {author} {\bibfnamefont {A.~B.}\ \bibnamefont {P\'erez-Cerd\'an}} \emph {et~al.},\ }\bibfield  {title} {\enquote {\bibinfo {title} {Deformation of {S}r and {R}b isotopes close to the ${N}={Z}$ line via $\ensuremath{\beta}$-decay studies using the total absorption technique},}\ }\href {\doibase 10.1103/PhysRevC.88.014324} {\bibfield  {journal} {\bibinfo  {journal} {Phys. Rev. C}\ }\textbf {\bibinfo {volume} {88}},\ \bibinfo {pages} {014324} (\bibinfo {year} {2013})}\BibitemShut {NoStop}%
\bibitem [{\citenamefont {Guadilla}\ \emph {et~al.}(2019)\citenamefont {Guadilla} \emph {et~al.}}]{Guadilla2019}%
  \BibitemOpen
  \bibfield  {author} {\bibinfo {author} {\bibfnamefont {V.}~\bibnamefont {Guadilla}} \emph {et~al.},\ }\bibfield  {title} {\enquote {\bibinfo {title} {Total absorption $\ensuremath{\gamma}$-ray spectroscopy of niobium isomers},}\ }\href {\doibase 10.1103/PhysRevC.100.024311} {\bibfield  {journal} {\bibinfo  {journal} {Phys. Rev. C}\ }\textbf {\bibinfo {volume} {100}},\ \bibinfo {pages} {024311} (\bibinfo {year} {2019})}\BibitemShut {NoStop}%
\bibitem [{\citenamefont {Dombos}\ \emph {et~al.}(2021)\citenamefont {Dombos} \emph {et~al.}}]{Dombos2021}%
  \BibitemOpen
  \bibfield  {author} {\bibinfo {author} {\bibfnamefont {A.~C.}\ \bibnamefont {Dombos}} \emph {et~al.},\ }\bibfield  {title} {\enquote {\bibinfo {title} {Total absorption spectroscopy of the $\ensuremath{\beta}$ decay of $^{101,102}\mathrm{Zr}$ and $^{109}\mathrm{Tc}$},}\ }\href {\doibase 10.1103/PhysRevC.103.025810} {\bibfield  {journal} {\bibinfo  {journal} {Phys. Rev. C}\ }\textbf {\bibinfo {volume} {103}},\ \bibinfo {pages} {025810} (\bibinfo {year} {2021})}\BibitemShut {NoStop}%
\bibitem [{\citenamefont {Mueller}\ \emph {et~al.}(2000)\citenamefont {Mueller} \emph {et~al.}}]{Mueller2000}%
  \BibitemOpen
  \bibfield  {author} {\bibinfo {author} {\bibfnamefont {W.~F.}\ \bibnamefont {Mueller}} \emph {et~al.},\ }\bibfield  {title} {\enquote {\bibinfo {title} {$\ensuremath{\beta}$ decay of ${}^{66}\mathrm{Co},$ ${}^{68}\mathrm{Co},$ and ${}^{70}\mathrm{Co}$},}\ }\href {\doibase 10.1103/PhysRevC.61.054308} {\bibfield  {journal} {\bibinfo  {journal} {Phys. Rev. C}\ }\textbf {\bibinfo {volume} {61}},\ \bibinfo {pages} {054308} (\bibinfo {year} {2000})}\BibitemShut {NoStop}%
\bibitem [{\citenamefont {Simon}\ \emph {et~al.}(2013)\citenamefont {Simon} \emph {et~al.}}]{Sim13}%
  \BibitemOpen
  \bibfield  {author} {\bibinfo {author} {\bibfnamefont {A.}~\bibnamefont {Simon}} \emph {et~al.},\ }\bibfield  {title} {\enquote {\bibinfo {title} {Su{N}: {S}umming {N}a{I}({T}l) gamma-ray detector for capture reaction measurements},}\ }\href {\doibase https://doi.org/10.1016/j.nima.2012.11.045} {\bibfield  {journal} {\bibinfo  {journal} {Nucl. Instr. Meth. Phys. Res. A}\ }\textbf {\bibinfo {volume} {703}},\ \bibinfo {pages} {16--21} (\bibinfo {year} {2013})}\BibitemShut {NoStop}%
\bibitem [{\citenamefont {Brown}\ and\ \citenamefont {Rae}(2014)}]{nushellx}%
  \BibitemOpen
  \bibfield  {author} {\bibinfo {author} {\bibfnamefont {B.A.}\ \bibnamefont {Brown}}\ and\ \bibinfo {author} {\bibfnamefont {W.D.M.}\ \bibnamefont {Rae}},\ }\bibfield  {title} {\enquote {\bibinfo {title} {The shell-model code {N}u{S}hell{X}@{M}{S}{U}},}\ }\href {\doibase https://doi.org/10.1016/j.nds.2014.07.022} {\bibfield  {journal} {\bibinfo  {journal} {Nucl. Data Sheets}\ }\textbf {\bibinfo {volume} {120}},\ \bibinfo {pages} {115--118} (\bibinfo {year} {2014})}\BibitemShut {NoStop}%
\bibitem [{\citenamefont {Chiara}\ \emph {et~al.}(2015)\citenamefont {Chiara} \emph {et~al.}}]{Chiara2015}%
  \BibitemOpen
  \bibfield  {author} {\bibinfo {author} {\bibfnamefont {C.~J.}\ \bibnamefont {Chiara}} \emph {et~al.},\ }\bibfield  {title} {\enquote {\bibinfo {title} {Identification of deformed intruder states in semi-magic $^{70}\mathrm{Ni}$},}\ }\href {\doibase 10.1103/PhysRevC.91.044309} {\bibfield  {journal} {\bibinfo  {journal} {Phys. Rev. C}\ }\textbf {\bibinfo {volume} {91}},\ \bibinfo {pages} {044309} (\bibinfo {year} {2015})}\BibitemShut {NoStop}%
\end{thebibliography}%
